\documentclass[5p,twocolumn]{elsarticle}

\usepackage{amsmath}    
\usepackage{amssymb}    
\usepackage{graphicx}   
\usepackage{hyperref}   
\usepackage{booktabs}   
\usepackage{fouriernc}
\journal{Automatica}

\usepackage{soul}  
\usepackage{array}
\usepackage{makecell}

\usepackage{amsmath,amssymb,amsfonts}
\usepackage{graphicx}
\usepackage{textcomp}
\usepackage{amssymb}
\usepackage{algorithm}
\usepackage{algorithmic}
\usepackage{subcaption}
\usepackage{multirow}
\usepackage{enumerate}
\usepackage{color}
\usepackage{amssymb}

\usepackage{amsthm}
\usepackage{setspace}
\usepackage{subcaption}
\usepackage{natbib}

\usepackage{nccmath}
\usepackage{hyperref}
\usepackage[dvipsnames]{xcolor}
\usepackage{wrapfig}
\usepackage{graphicx}
\usepackage{fancyhdr}
\floatname{algorithm}{Algorithm}

\newtheoremstyle{definition}
  {\topsep}   
  {\topsep}   
  {\normalfont}  
  {}           
  {\bfseries}  
  {.}          
  {.5em}       
  {}           
\theoremstyle{definition}
\newtheorem{definition}{Definition}

\newtheorem{example}{Example}

\newtheorem{lemma}{Lemma}
\newtheorem{assumption}{Assumption}

\newtheorem{theorem}{Theorem}
\theoremstyle{plain}

\begin{document}

\begin{frontmatter}
\title{A Distributed Actor-Critic Algorithm for Fixed-Time Consensus in Nonlinear Multi-Agent Systems}

\author[inst1]{Aria Delshad}
\ead{aria.delshad@ee.sharif.edu}
\author[inst1]{Maryam Babazadeh\corref{cor1}}
\ead{babazadeh@sharif.edu}

\cortext[cor1]{Corresponding author.}

\address[inst1]{Department of Electrical Engineering, Sharif University of Technology, Tehran, Iran}

\begin{abstract}
This paper proposes a reinforcement learning (RL)-based backstepping control strategy to achieve fixed time consensus in nonlinear multi-agent systems with strict feedback dynamics. Agents exchange only output information with their neighbors over a directed communication graph, without requiring full state measurements or symmetric communication.
Achieving fixed time consensus, where convergence occurs within a pre-specified time bound that is independent of initial conditions is faced with significant challenges due to the presence of unknown nonlinearities, inter-agent couplings, and external disturbances. This work addresses these challenges by integrating actor critic reinforcement learning with a novel fixed time adaptation mechanism. Each agent employs an actor critic architecture supported by two estimator networks designed to handle system uncertainties and unknown perturbations.
The adaptation laws are developed to ensure that all agents track the leader within a fixed time regardless of their initial conditions. The consensus and tracking errors are guaranteed to converge to a small neighborhood of the origin, with the convergence radius adjustable through control parameters. Simulation results demonstrate the effectiveness of the proposed approach and highlight its advantages over state-of-the-art methods in terms of convergence speed and robustness.
\end{abstract}

\begin{keyword}
Multi-agent Systems \sep Reinforcement Learning \sep Fixed-time Consensus \sep Distributed Control \sep Backstepping Control
\end{keyword}

\end{frontmatter}

\section{Introduction}\label{sec:introduction}
Over the past two decades, cooperative control in multi-agent systems has gained significant research attention due to its scalability and robustness in executing group tasks. While most research has focused on systems with first and second-order linear or nonlinear dynamics, many practical engineering applications require higher-order models \cite{Amirkhani2022-fy}, \cite{5409635}.

The strict-feedback structure has become a foundational framework for the control of high-order nonlinear systems, particularly in multi-agent scenarios \cite{Wen2012-ut}. While backstepping methods are widely used for such systems due to their recursive design capability, the incorporation of optimal control principles remains underexplored. This gap primarily stems from the computational and theoretical challenges involved in solving the Hamilton-Jacobi-Bellman (HJB) equation in the context of interconnected agents \cite{Zhou2022-jv}.

Reinforcement learning (RL) offers a promising alternative by providing approximate solutions to optimal control problems without explicitly solving the HJB equation \cite{Sutton2018-bv}.  Accordingly, the integration of RL with backstepping through a neural network-based actor-critic architecture is introduced in \cite{Wen2018-nt}. However, the method in \cite{Wen2018-nt} assumes known system dynamics and does not readily extend to multi-agent environments.  Subsequent research has aimed to extend RL-based backstepping approaches to more complex and practical scenarios. For instance, in the case of known multi-agent systems in integral-based strict-feedback form, \cite{9906584} proposes an integrated framework that combines sliding mode control with reinforcement learning. This approach successfully leads to the development of composite distributed control protocols for multi-agent systems. For second-order multi-agent systems with unknown dynamics in strict-feedback form, \cite{9635597} presents an actor-critic approach that achieves semi-global uniformly ultimately bounded (SGUUB) consensus. This method, however, does not account for external disturbances and assumes an undirected communication graph. Moreover, the system nonlinearities are presumed to be bounded. This framework is extended in \cite{9525047} to accommodate arbitrary-order agents, still within the same strict-feedback structure and under unknown dynamics. In \cite{9086141}, the RL-based backstepping technique is further adapted for fault-tolerant tracking control by incorporating adaptive mechanisms.
To improve communication efficiency and enhance robustness, \cite{9470911} proposes an event-triggered RL control strategy that is resilient to sensor faults. A different perspective is introduced in \cite{9787792}, where a game-theoretic backstepping method solves the Hamilton–Jacobi–Isaacs (HJI) equation approximately via policy iteration for each subsystem. A single critic neural network is used to reduce computational complexity, although the method relies on the persistency of excitation (PE) condition for convergence.
Recent efforts have also prioritized transient and steady-state performance. In \cite{YAN2022649}, a finite-time scaling function is integrated into the RL-based backstepping design for bidirectional communication networks, achieving faster convergence and improved tracking accuracy, albeit still requiring the PE condition. Meanwhile, \cite{10286389} considers a simplified strict-feedback structure in which uncertainties and nonlinearities are restricted to the final layer, with preceding dynamics modeled as pure integrator chains. The proposed method combines adaptive neural networks, a disturbance observer, and experience replay to ensure uniformly ultimately bounded (UUB) performance under bounded disturbances affecting the final subsystem. Containment control for unknown multi-agent systems in strict-feedback form is explored in \cite{Luo2022-ah}, offering a strategy that constrains consensus errors within desired bounds. However, this method does not explicitly address robustness against external disturbances.

Recent studies have increasingly focused on finite-time and fixed-time stabilization to achieve faster convergence and improved control performance in multi-agent systems \cite{9696292}. To cope with system uncertainties and nonlinear dynamics, \cite{9158531} proposes a distributed adaptive control framework that guarantees finite-time stability for stochastic strict-feedback multi-agent systems. For second-order agent dynamics, \cite{9007508} addresses the fixed-time consensus problem under output constraints and unknown control directions, providing a robust leader-follower coordination strategy. The communication topology in this work is undirected, and external disturbances are only considered at the second layer. Containment control within a fixed-time framework is explored in \cite{Cui2022-fv}, which develops a fault-tolerant consensus protocol using a network-based actuator fault observer. The proposed control design ensures convergence to the convex hull defined by dynamic leaders.
Additionally, \cite{XU2022127176} presents a finite-time optimal consensus strategy that leverages customized barrier functions to manage time-varying and asymmetric output constraints in systems with unknown follower dynamics. However, it assumes that the leader's dynamics are fully known and that disturbance signals are energy-bounded. To enhance robustness against general disturbances while ensuring fixed-time convergence, \cite{10309212} introduces an identifier-actor-critic architecture alongside a novel quadratic cost function. The method targets second-order agents and achieves fixed-time consensus in the presence of disturbances affecting the final layer.

This paper introduces a novel distributed RL framework for achieving fixed-time consensus in leader-follower multi-agent systems with agents of arbitrary order. Each agent is modeled in strict-feedback form, subject to unknown nonlinear dynamics and external bounded disturbances. The proposed control architecture employs optimized backstepping in combination with actor-critic neural networks, enhanced by newly designed fixed-time adaptation laws. Achieving fixed-time consensus—where convergence is guaranteed within a pre-specified time bound independent of initial conditions— is faced by significant challenges. These include managing complex nonlinearities, inter-agent couplings, and system uncertainties without relying on restrictive assumptions such as bounded dynamics or the Persistency of Excitation condition. This work addresses these challenges through careful controller and adaptation law design. The main contributions can be summarized as follows:
\begin{itemize}
    \item A novel distributed RL-based control algorithm is proposed to achieve fixed-time consensus in multi-agent systems via optimized backstepping.
    \item The proposed method guarantees that the consensus error converges to a neighborhood of the origin within a fixed time. The neighborhood radius is tunable via control parameters.
    \item   New fixed-time adaptation laws are designed for the actor, critic, and observer networks. These ensure convergence without requiring the PE condition, which is typically needed for parameter convergence in RL-based schemes.
    \item In contrast to the relevant existing studies \cite{9635597, 9525047, YAN2022649, 10286389, 9007508, 10309212}, the proposed approach accommodates both directed and undirected communication graphs. It does not require the unknown nonlinearities to be bounded, and allows nonlinear uncertainties to appear in every layer of the strict-feedback dynamics.
    \item Robustness to bounded disturbances is ensured across all layers of the agent dynamics, thereby generalizing prior works in which robustness is either not considered \cite{9635597, 9525047, 9787792, YAN2022649, Luo2022-ah}, limited to disturbances affecting only the last layer \cite{10286389, 9007508, 10309212}, or restricted to bounded-energy disturbances \cite{XU2022127176}.
\end{itemize}

\section{Preliminaries}
\subsection{Fixed-time Stability}
Consider a nonlinear system represented by,
 \begin{equation}\label{eq:nonlinear1}
  \dot{x}(t)=f(x(t)),\qquad x(0)=x_0,
 \end{equation}
where $x(t) \in \mathbb{R}^n, f(x(t)): \mathbb{R}^n \to \mathbb{R}^n$ is a continuous function. Assuming a unique solution exists for each initial state in \eqref{eq:nonlinear1}, consider the following definitions:
\begin{definition} (\cite{doi:10.1137/S0363012997321358})
The equilibrium point of \eqref{eq:nonlinear1} is globally finite-time stable if it meets the criteria of global asymptotic stability and ensures that all system responses converge to this point within a finite time.
\end{definition}

\begin{definition} (\cite{6104367})
The equilibrium point of \eqref{eq:nonlinear1} is considered fixed-time stable if it is globally finite-time stable and its settling time $T(x_0)$
is finite such that $T(x_0) < T_{\max}$ for $ T_{\max} > 0$. This indicates that the settling time's upper bound is independent of the system's initial state.
\end{definition}

\begin{lemma} (\cite{Ni2016-cx})
Consider the system \eqref{eq:nonlinear1} with a positive continuous function \( V(x) \) such that,
 \begin{equation}
  \dot{V}(x) \leq -k_p V^p - k_q V^q + C,
 \end{equation}
where, $k_p$, $k_q$, $p$, $q$ and $C$ are all positive numbers, with \( p < 1 \) and \( q > 1 \) being ratios of odd numbers. Then the equilibrium point of system (1) is fixed-time stable. The region of convergence is given by,

 \begin{equation} \label{eq:Omega}
  \Omega=\left\{x \left\lvert\, V(x) \leq \min \left\{\left(\frac{C}{(1-\vartheta) k_p}\right)^{\frac{1}{p}},\left(\frac{C}{(1-\vartheta) k_q}\right)^{\frac{1}{q}}\right\}\right.\right\}
 \end{equation}
which $0<\vartheta<1$ is a constant. The upper bound of the settling time $T_{\max}$ is given by $ T_{\max } \leq \frac{1}{k_p(1-p)}+\frac{1}{k_q(q-1)}$.
\end{lemma}

\begin{lemma} (\cite{Ni2016-cx,doi:10.1080/00207179.2013.834484})
    If $V_j$, $j=1, 2, \cdots, n$ are non-negative and  $V=\sum_{j=1}^n V_j, j=1,2, \ldots, n$, then,
 \begin{equation}
    \left\{
        \begin{aligned} 
        &   \sum_{i=1}^n V_j{ }^p \geq V^p   &   0 < p \leq 1\\
        &  \sum_{i=1}^n V_j{ }^p \geq n^{1-p} V^p & p>1.
    \end{aligned}
    \right.
\end{equation}
\end{lemma}

\begin{lemma} (\cite{Ni2016-cx,Hardy1988-lt})
(Young's inequality): For any $a, b \in \mathbb{R}$, and positive constants $p, q, c$ such that $p, q > 1$ and $\frac{1}{p} + \frac{1}{q} = 1$, the following inequality holds,
\begin{equation}
    a b \leq \frac{c^p}{p}|a|^p+\frac{1}{c^q q}|b|^q,
\end{equation}
\end{lemma}
In the following, two algebraic lemmas are presented to facilitate the fixed-time stability analysis. 
\begin{lemma}
   Let the vectors $a,b \in \mathbb{R}^n$ be given as ${a=\left[a_1, a_2, \cdots, a_n\right]^\top, b=\left[b_1, b_2, \cdots, b_n\right]^\top}$. Then, 
    \begin{equation}
        -a^\top(a+b)^{\frac{1}{3}} \leq -\frac{1}{2}\sum_{i=1}^n\ a_i^{\frac{4}{3}}+ \sum_{i=1}^n b_i^{\frac{4}{3}}.
    \end{equation}
\end{lemma}
\noindent\textbf{Proof.} The proof is given in Appendix A.

\begin{lemma}
    For two vectors $a,b \in \mathbb{R}^n$,
    \begin{equation}
    -a^\top(a+b)^3 \leq -\frac{1}{8} \sum_{i=1}^n a_i^4 + 172 \sum_{i=1}^n b_i^4.
\end{equation}
\end{lemma}
\noindent\textbf{Proof.} The proof is given in Appendix B.

\subsection{Optimal Control Formulation}
Consider the nonlinear system,
\begin{equation}\label{eq:nonlinear2}
\dot{x}(t)=f(x(t), u(x(t))), 
\end{equation}
where $x(t) \in \mathbb{R}^n$ is the state variable, $u(x(t)) \in \mathbb{R}^m$ is the control input, and $f(x(t), u(x(t))) \in \mathbb{R}^n$ is a continuous function with $f(0)=0$. Consider the integral performance index over an infinite horizon given by,
\begin{equation}\label{eq:index}
    J(x, u)=\int_t^{\infty} c(x(\tau), u(x(\tau))) d \tau.
\end{equation}
Here, $c(x, u) = x^\top P_1 x + u^\top(x) P_2 u(x)$ is the cost function. The matrices $P_1 = P_1^\top \in\mathbb{R}^{n\times n}$ and $P_2 = P_2^\top \in \mathbb{R}^{m\times m}$ are positive semi-definite and positive definite, respectively.

For the system represented by \eqref{eq:nonlinear2} and the value function given by \eqref{eq:index}, the HJB equation is defined as follows, \cite{Rizvi2022-hu},
\begin{equation}\label{eq:HJB}
  \min _{u \in U}\left(c(x, u)+\left(\frac{J(x, u)}{\partial x}\right)^\top f(x, u)\right)=0 . 
\end{equation}
The set $U$ denotes the set of admissible control inputs. The solution to the HJB equation yields the optimal value function ${J}^*$. The optimal control policy can be determined by finding the value of $u^*(x,t)$ that minimizes the expression inside the minimum operator in the HJB equation \eqref{eq:HJB} \cite{Rizvi2022-hu}, that is,

\begin{equation}
    u^*(x, t)=\arg \min _{u \in U}\left(c(x, u)+\left(\frac{J(x, u)}{\partial x}\right)^\top f(x, u)\right).
\end{equation}

Now, let the dynamics of system \eqref{eq:nonlinear2} be given as,
\begin{equation}\label{eq:nonlinear3}
    \dot{x}(t)=f(x)+g(x) u(x).
\end{equation}


    



Then the HJB equation is derived as,
\begin{equation}\label{eq:11}
 \begin{array}{r}
     H\left(x, u^*, J^*\right)=x^\top P_1 x+u^{*^\top}(x) P_2 u^*(x)  \\
          +\left(\frac{J^*(x, u)}{\partial x}\right)^\top \times\left(f(x)+g(x) u^*(x)\right)=0,
 \end{array}
\end{equation}
and $u^*(x)$ can be derived from the solution $\partial H\left(x, u^*, J^*\right) / \partial u^*=0$ as follows,

 \begin{equation}\label{eq:12}
    u^*(x)=-\frac{1}{2} P_2^{-1} g^\top(x) \frac{d J^*(x, u)}{d x}.
 \end{equation}
To derive the optimal control policy and to ensure \eqref{eq:12} as the unique control solution, consider the following HJB equation, which is derived by substituting \eqref{eq:12} into \eqref{eq:11},
\begin{equation}\label{eq:13}
 \begin{array}{r}
H\left(x, u^*, J^*\right)=x^\top P_1 x+\frac{d J^*(x)}{d x^\top} f(x) 
\\
-\frac{1}{4} \frac{d J^*(x)}{d x^\top} g(x)\times P_2^{-1} g^\top(x) \frac{d J^*(x)}{d x}=0.
 \end{array}
\end{equation}
Finding the analytical solution of the HJB equation \eqref{eq:13} is challenging. To overcome this challenge, RL is employed as a practical method to approximate the optimal control solution \cite{Rizvi2022-hu}. Among RL methods, actor-critic algorithms are widely applied in optimal control and form the foundation for the distributed algorithm proposed in this paper.

 \subsection{Topology of communication graph}
The communication graph’s topology plays a pivotal role in shaping the overall performance of a multi-agent system. In this study, the communication graph can be either directed or undirected and is represented as $\mathcal{G}=(\mathcal{V}, \mathcal{E}, A)$, where $V=\left[v_1, v_2, \cdots, v_n\right]$ denotes the set of nodes (agents), $\mathcal{E} \subseteq \mathcal{V} \times \mathcal{V}$ signifies the set of edges, and ${A} = [{a}_{ik}] \in \mathbb{R}^{n\times n}$ is the adjacency matrix.

\begin{assumption}
The directed graphs are assumed to include a spanning tree with the leader node as the root. In this case, at least one agent is connected to the leader, ensuring that $b_1 + b_2 + \cdots + b_n > 0$ where $b_1,b_2,\cdots,b_n$ are the communication weights between the followers and leader. 
\end{assumption}

 \begin{lemma} (\cite{POLYAKOV2015332})
Under the condition of Assumption 1, $\widetilde{L}=L+B$  is recognized as an invertible matrix, where $B=\text{diag}(b_1,b_2,\cdots,b_n)$. This property ensures that the system is controllable and that consensus is achievable in the presence of a leader.
 \end{lemma}

\begin{lemma} (\cite{Wen2018-nt})
    The Laplacian matrix of an undirected connected graph is irreducible.
\end{lemma}

 \begin{lemma} (\cite{Wen2018-nt})
If the Laplacian matrix $L$ is irreducible, then the matrix $\widetilde{L}=L+B$ is a positive definite matrix, where $B > 0$, which means that $\widetilde{L}$ is invertible.
 \end{lemma}

\section{Problem Formulation}\label{sec:3}
Consider a multi-agent system with $N$ followers and one leader. The dynamics of the leader is given in strict-feedback form as,

\begin{equation}\label{eq:leader}
    \left\{\begin{array}{l}
\dot{x}_{01}(t)=x_{02}(t)+f_{01}\left(x_{01}(t)\right)+d_{01}(t) \\
\dot{x}_{02}(t)=x_{03}(t)+f_{02}\left(\bar{x}_{02}(t)\right)+d_{02}(t) \\
\vdots \\
\dot{x}_{0 n}(t)=u_0(t)+f_{0 n}\left(\bar{x}_{0 n}(t)\right)+d_{0 n}(t) \\
y_0(t)=x_{01}(t),
\end{array}\right.
\end{equation}
Where $\bar{x}_{0k} = [x_{01}, x_{02}, \cdots, x_{0k}]^\top$, $k=1,2,\cdots,n$. ${x}_{0k}\in\mathbb{R}$ is the $k$th state variable of the leader agent, $u_0 (t) \in \mathbb{R}$ is the
control input of the leader, $y_0(t) \in \mathbb{R}$ represents the output of the leader; ${f}_{0k}(\bar{x}_{0k}(t)) : \mathbb{R}^{{n}_{0k}}\to \mathbb{R}$ is a smooth function, and ${d}_{0k}$ represents the external disturbance to the $k$th state variable of the leader. It is assumed that  $u_0 (t) = k_0 (x_{01}(t), x_{02}(t), \cdots, x_{0n}(t), y_{0d}(t), \dot{y}_{0d}(t), \cdots, y_{0d}^{(n)}(t)) $ exists which forces the leader agent to follow the desired
path $y_{od} (t)$. Similarly, the dynamics of the followers are given as,

\begin{equation}\label{eq:followers}
    \left\{\begin{array}{l}
\dot{x}_{i 1}(t)=x_{i 2}(t)+f_{i 1}\left(x_{i 1}(t)\right)+d_{i 1}(t) \\
\dot{x}_{i 2}(t)=x_{i 3}(t)+f_{i 2}\left(\bar{x}_{i 2}(t)\right)+d_{i 2}(t) \\
\vdots \\
\dot{x}_{i n}(t)=u_i(t)+f_{i n}\left(\bar{x}_{i n}(t)\right)+d_{i n}(t) \\
y_i(t)=x_{i 1}(t),
\end{array}\right.
\end{equation}
Where $\bar{x}_{ik} = [x_{i1}, x_{i2}, \cdots, x_{ik}]^\top$, $k=1,2,\cdots,n$ and $i=1,2,\cdots,N$. Here, $N$ is the number of followers and $n$ is the number of state variables of each agent. ${x}_{ik} \in \mathbb{R}$ is the $k$th state variable of the $i$th follower, $u_i (t) \in \mathbb{R}$ is the control input
of the $i$th follower, $y_i(t) \in\mathbb{R}$ is the output of the $i$th follower, ${f}_{ik}(\bar{x}_{ik}(t)) : \mathbb{R}^{{n}_{ik}}\to \mathbb{R}$ is a smooth function, and ${d}_{ik} (t)$  represent the external disturbance to the $k$th state variable of the $i$th follower. 
 \begin{assumption}
The dynamics of the leader and its control input, i.e. ${f}_{0k}(\bar{x}_{0k}(t))$ and $u_0 (t)$, are unknown to the followers.
 \end{assumption}

 \begin{assumption}
External disturbances and their derivatives for both the leader and the follower are bounded, i.e. $ \left|d_{i k}(t)\right| <\bar{d}_{i k},\qquad\left|\dot{d}_{i k}(t)\right|<\bar{d}_{d, i k}$ for $k=1,2,\cdots,n$, and ${i=0,1, \cdots, N}$ 
 \end{assumption}
The objective is to develop a distributed backstepping controller, based on reinforcement learning, for the followers \eqref{eq:followers} under assumptions 1 to 3 to track the leader \eqref{eq:leader}. The control design is required to ensure that the follower outputs $y_i (t)$ track the leader output $y_0 (t)$ within a fixed-time frame, with an upper bound that is independent of the system’s initial conditions.

\section{Distributed Control Design based on RL}
The tracking error between the leader and the $i$th follower is defined as,
 \begin{equation}
     z_{i1}(t) = y_i(t) - y_0(t).
 \end{equation}
Let the virtual errors for the backstepping control design be defined as,

\begin{equation}
    z_{ij}(t) = x_{ij}(t) - \alpha_{i(j-1)}(t), \qquad j=2, \cdots, n
\end{equation}
Where $\alpha_{i(j-1)}(t)$ represents the $(j-1)$th virtual control law for the $i$th follower. Moreover, the consensus error is,
\begin{equation}
    e_i(t)=\sum_{l \in N_i} a_{i l}\left(y_i(t)-y_l(t)\right)+b_i\left(y_i(t)-y_0(t)\right).
\end{equation}
Here, $N_i$ denotes the set of neighbors of the $i$th agent in the communication graph. The tracking error and the consensus error are related to each other by,
\begin{equation}
    e(t) = \tilde{L} z_1(t),
\end{equation}
Where $\tilde{L}=L+B$, $z_1 (t) = [z_{11} (t), z_{21} (t), \cdots, z_{N1} (t)]^\top$, and $e(t) = [e_1 (t), e_2 (t), \cdots, e_N (t)]^\top$. Based on Lemmas 6, 7, and 8, the matrix $\tilde{L}$ is invertible. Consequently, as $e(t)$ approaches zero, $z_1(t)$ will also converge to zero.

The proposed control scheme in this paper involves breaking down the dynamics of the followers into a set of subsystems and progressively designing control policies for each step. Starting with the innermost subsystem in the strict feedback form, a control policy is designed as step 1. The details of this design are elaborated in Section 4.1. The policy is then incorporated into the next outer layer of the system, and the process is repeated iteratively until the entire system achieves fixed-time consensus. The key idea is to ensure that each step's stability contributes to the overall stability of the entire system, effectively. Accordingly, in Section 4.2, the developed virtual policies for steps $j=2,\ldots n-1$, and the actual control policy for step $n$ are derived.\\

\subsection{Step 1 for RL-based backstepping}
The consensus error can be represented as,
 \begin{equation}
 \begin{aligned}
      e_i(t)&=\left(\sum_{l \in N_i} a_{i l}+b_i\right) y_i(t)-\sum_{l \in N_i} a_{i l} y_l(t)-b_i y_0(t) \\
      & =g_i y_i(t)-\sum_{l \in N_i} a_{i l} y_l(t)-b_i y_0(t),
 \end{aligned}
 \end{equation}
and the time derivative of $e_i (t)$ is,
\begin{equation}
    \dot{e}_i(t)=g_i \dot{y}_i(t)-\sum_{l \in N_i} a_{i l} \dot{y}_l(t)-b_i \dot{y}_0(t).
\end{equation}
Based on the dynamics of the leader and followers,

\begin{equation}
    \begin{aligned}
       &\dot{e}_i(t)=g_i\left(x_{i 2}(t)\right. \left.+f_{i 1}\left(x_{i 1}(t)\right)+d_{i 1}(t)\right) \\
       & -\sum_{l \in N_i}  a_{i l}\left(x_{l 2}(t)+f_{l 1}\left(x_{l 1}(t)\right)+d_{l 1}(t)\right) \\
       & -b_i\left(x_{02}(t)+f_{01}\left(x_{01}(t)\right)+d_{01}(t)\right).
    \end{aligned}
\end{equation}
Thus, the consensus error dynamics can be written in a compact form as,

\begin{equation}
    \dot{e}_i(t)=g_i x_{i 2}(t)+F_{i 1}\left(X_{i 1}(t)\right)+D_{i 1}(t),
\end{equation}
with $X_{i1}(t) = [x_{i1}(t), \bar{x}^\top_{l2}, \bar{x}^\top_{02}]^\top$ and the unknown terms $F_{i 1}(X_{i 1}(t))$ and $D_{i 1}(t)$ be given as,
\begin{equation}
 \begin{aligned}
      F_{i 1}\left(X_{i 1}(t)\right) &=g_i f_{i 1}\left(x_{i 1}(t)\right) -\sum_{l \in N_i} a_{i l}\left(x_{l 2}(t)+f_{l 1}\left(x_{l 1}(t)\right)\right) \\
     & -b_i\left(x_{02}(t)+f_{01}\left(x_{01}(t)\right)\right),\\
      D_{i 1}(t) &=g_i d_{i 1}(t)-\sum_{l \in N_i} a_{i l} d_{l 1}(t)-b_i d_{01}(t).
 \end{aligned}
\end{equation}
The virtual control policy $\alpha_{i1}(t)$ of the first step is composed of two components. The optimal control policy $v_{i1}^*(t)$ and the fixed -time control policy $h_{i1}(t)$. That is,

\begin{equation}\label{eq:virtual1}
    \alpha_{i1}(t)=v_{i1}^*(t)+h_{i1}(t).
\end{equation}

To design the optimal control law in the first step, let the performance index function be described as,
\begin{equation}\label{eq:Jindex}
    J_{i 1}\left(e_i(t)\right)=\int_t^{+\infty}\left(e_i^2(\tau)+v_{i 1}^2(\tau)\right) d \tau.
\end{equation}
Assuming that $v_{i1}^*(t)$ is the minimizer of  \eqref{eq:Jindex}, the optimal performance index is, 
\begin{equation}
    \begin{gathered}
J_{i 1}^*\left(e_i(t)\right)=\int_t^{+\infty}\left(e_i^2(\tau)+v_{i 1}^*{ }^2(\tau)\right) d \tau,
\end{gathered}
\end{equation}
\begin{equation}
   \frac{d J_{i 1}^*\left(e_i(t)\right)}{d t}=-\left(e_i^2(t)+{v_{i 1}^*}^2(t)\right).
\end{equation}
The resulting HJB equation is, 
\begin{equation}
   \begin{aligned}
     & H_{i 1}\left(e_i(t), v_{i 1}^*(t), J_{i 1}^*\right)=e_i^2(t)+{v_{i 1}^*}^2 (t)+  \\
     & \frac{d J_{i 1}^*\left(e_i(t)\right)}{d e_i(t)}\left(g_i v_{i 1}^* (t)+F_{i 1}\left(X_{i 1}(t)\right)+D_{i 1}(t)\right) .
   \end{aligned}
\end{equation}
Accordingly, $v_{i1}^*(t)$ can be obtained by solving,
\begin{equation}
    \frac{\partial H_{i 1}\left(e_i(t), v_{i 1}^*(t), J_{i 1}^*\left(e_i(t)\right)\right)}{\partial v_{i 1}^*(t)}=0,
\end{equation}
or equivalently,
\begin{equation}
    2 v_{i 1}^*(t)+g_i \frac{d J_{i 1}^*\left(e_i(t)\right)}{d e_i(t)}=0.
\end{equation}
Thus, the optimal control law of the first step is equal to,
\begin{equation}
    v_{i 1}^*(t)=-\frac{1}{2} g_i \frac{d J_{i 1}^*\left(e_i(t)\right)}{d e_i(t)}.
\end{equation}
Note that $\frac{d J_{i 1}^*\left(e_i(t)\right)}{d e_i(t)}$ is not available. By introducing the new term $J_{i 1}^r\left(e_i(t)\right)$ as,
\begin{equation}
    J_{i 1}^r\left(e_i(t)\right)=-k_{i 1} e_i(t)+\frac{g_i{ }^2}{2} \frac{d J_{i 1}^*\left(e_i(t)\right)}{d e_i(t)},
\end{equation}
the control signal $v_{i 1}^*(t)$ can be represented as,
\begin{equation}
    v_{i 1}^*(t)=-\frac{k_{i 1}}{g_i} e_i(t)-\frac{1}{g_i} J_{i 1}^r\left(e_i(t)\right),
\end{equation}
where $J_{i 1}^r$ is the new unknown term to be approximated by RL and $k_{i 1}$ is an arbitrary positive constant.
Next, $J_{i 1}^r$ is approximated using a neural
network as,
\begin{equation}
    J_{i 1}^r\left(e_i(t)\right)=W_{i 1}^{{*}^\top} S_{i 1}\left(e_i(t)\right)+\varepsilon_{i 1}\left(e_i(t)\right),
\end{equation}
where $W_{i 1}^{*}$ is the vector of ideal weights, $S_{i1}$ are the activation functions, and $\varepsilon_{i 1}$ is the estimation error of the neural network related to RL in step 1. 
Accordingly,
\begin{equation}
    v_{i 1}^*(t)=-\frac{k_{i 1}}{g_i} e_i(t)-\frac{1}{g_i} W_{i 1}^{{*}^\top} S_{i 1}\left(e_i(t)\right)-\frac{1}{g_i} \varepsilon_{i 1}\left(e_i(t)\right),
\end{equation}
\begin{equation}
    \frac{d J_{i 1}^*\left(e_i(t)\right)}{d e_i(t)}=\frac{2 k_{i 1}}{g_i^2} e_i(t)+\frac{2}{g_i^2} W_{i 1}^{*^\top} S_{i 1}\left(e_i(t)\right)+\frac{2}{g_i^2} \varepsilon_{i 1}\left(e_i(t)\right).
\end{equation}
 Given that the weights of $W_{i 1}^*$ are unknown, we will utilize their
estimation. Consequently, the derivative of the approximated optimal value function of the first step is,
\begin{equation}
    \frac{d \hat{J}^*_{i 1}\left(e_i(t)\right)}{d e_i(t)}=\frac{2 k_{i 1}}{g_i^2} e_i(t)+\frac{2}{g_i^2} \underbrace{\hat{W}^\top_{c, i 1}(t) S_{i 1}\left(e_i(t)\right).}_{\text {Critic } N N}
\end{equation}
Thus, the approximated optimal virtual control law corresponding to the first step is,
\begin{equation}
    \hat{v}_{i 1}^*(t)=-\frac{k_{i 1}}{g_i} e_i(t)-\frac{1}{g_i} \underbrace{\hat{W}^\top_{a, i 1}(t) S_{i 1}\left(e_i(t)\right).}_{\text {Actor } N N}
\end{equation}
Consider the following update rules for the weights of the actor and critic neural networks, respectively, where $\sigma_{1c,i1}, \sigma_{2c,i1}, \sigma_{3c,i1}, \sigma_{1a,i1}, \sigma_{2a,i1}, \sigma_{3a,i1} > 0$,
$\Gamma_{c, i 1}, \Gamma_{a, i 1} \succ 0$, and $p=\frac{1}{3}$, $q=3$.

\begin{equation}
\left\{
\begin{aligned}
    &\dot{\hat{W}}_{c,i1}(t) = \Gamma_{c,i1} \bigg( -S_{i1}\left(e_i(t)\right) e_i(t) \\
    & -\sigma_{1c,i1} S_{i1}\left(e_i(t)\right) S^\top_{i1}\left(e_i(t)\right) \hat{W}_{c,i1}(t) \\
    & -\sigma_{2c,i1} \hat{W}_{c,i1}^{{p}}(t) -\sigma_{3c,i1} \hat{W}_{c,i1}^q(t) \bigg),  \\
    &\dot{\hat{W}}_{a,i1}(t) =\\
    &\underbrace{-\Gamma_{a,i1} \sigma_{1a,i1} S_{i1}\left(e_i(t)\right) S_{i1}^\top\left(e_i(t)\right)\left(\hat{W}_{a,i1}(t) - \hat{W}_{c,i1}(t)\right)}_{A_{w a,i1}^1} \\
    & \underbrace{-\Gamma_{a,i1} \sigma_{2a,i1} \left(\hat{W}_{a,i1}(t) - \hat{W}_{c,i1}(t)\right)^p}_{A_{w a,i1}^2} \\
    & \underbrace{-\Gamma_{a,i1} \sigma_{3a,i1} \left(\hat{W}_{a,i1}(t) - \hat{W}_{c,i1}(t)\right)^q}_{A_{w a,i1}^3}, \label{eq:actor_critic1}
\end{aligned}
\right.
\end{equation}
In the actor adaptation, the term $A_{w a, i 1}^1$ enforces satisfaction of the HJB, while the terms $A_{w a, i 1}^2$ and $A_{w a, i 1}^3$ are included to ensure stability of the system in fixed time. 
Substituting ${\hat{J}^*_{i 1}\left(e_i(t)\right)}$ and $\hat{v}_{i 1}^*(t)$ in the HJB equation yields,
\begin{equation}
    \begin{aligned}
& H_{i 1}\left(e_i(t), \hat{v}_{i 1}^*(t), \hat{J}_{i 1}^*(e_i(t))\right)= \\
&e_i^2(t)+\left(-\frac{k_{i 1}}{g_i} e_i(t)-\frac{1}{g_i} \hat{W}_{a, i 1}^\top(t) S_{i 1}\left(e_i(t)\right)\right)^2 \\
&+\left(\frac{2 k_{i 1}}{g_i^2} e_i(t)+\frac{2}{g_i^2} \hat{W}_{c, i 1}^\top(t) S_{i 1}\left(e_i(t)\right)\right) \\
&\left(g_i\left(-\frac{k_{i 1}}{g_i}-\frac{1}{g_i} \hat{W}_{a, i 1}^\top(t) S_{i 1}\left(e_i(t)\right)\right)\right. \left.+F_{i 1}\left(X_{i 1}(t)\right)+D_{i 1}(t)\right).
\end{aligned}
\end{equation}
The optimized consensus control input \( \hat{v}_{i1}^*(t) \) is expected to be the unique solution of the equation  
$H_{i1}\left(e_i(t), \hat{v}_{i1}^*(t), \hat{J}_{i1}^*(e_i(t))\right) = 0.$
If this equation is satisfied and admits a unique solution, then,
\begin{equation}\label{eq:derivitive1}
\begin{aligned}
    &\frac{\partial H_{i 1}\left(e_i(t), \hat{v}^*_{i 1}(t), \hat{J}^*_{i 1}(e_{i}(t))\right)}{\partial \hat{W}_{a, i 1}(t)} = \\
    & \frac{2}{g_i^2} S_{i 1}\left(e_i(t)\right) S_{i 1}^\top\left(e_i(t)\right)\left(\hat{W}_{a, i 1}(t)-\hat{W}_{c, i 1}(t)\right)=0.
\end{aligned}
\end{equation}
Given the matrix,
\begin{small}
 \begin{equation}
    P(t)=\left(\hat{W}_{a, i 1}(t)-\hat{W}_{c, i 1}(t)\right)^\top S_{i 1}\left(e_i(t)\right)S_{i 1}^\top\left(e_i(t)\right)\left(\hat{W}_{a, i 1}(t)-\hat{W}_{c, i 1}(t)\right),
\end{equation}   
\end{small}
it is straightforward to see 
$P(t)=0$  is equivalent to \eqref{eq:derivitive1}.
The adaptation rule of the actor weights tries to minimize $P(t)$ by changing in the direction opposite to the gradient of $P(t)$, that is,
\begin{equation}
\begin{aligned}
  &  \dot{\hat{W}}_{a, i 1}(t)=-\Gamma_{a, i 1} \sigma_{1 a, i 1} \frac{\partial P(t)}{\partial \hat{W}_{a, i 1}(t)} \\
    & = -\Gamma_{a, i 1} \sigma_{1 a, i 1} S_{i 1}\left(e_i(t)\right) S_{i1}^\top\left(e_i(t)\right)\left(\hat{W}_{a, i 1}(t)-\hat{W}_{c, i 1}(t)\right),
\end{aligned}
\end{equation}
where the adaptation rule exactly meets  $A_{w a, i 1}^1$ in \eqref{eq:actor_critic1}. The next step is to devise the structure of the fixed-time stabilizer term and its associated adaptation laws in the context of the first step’s virtual control law. Note that by incorporation of the actor and critic weights, the dynamics of the consensus error is expressed as,
\begin{align}
  \nonumber  & \dot{e}_i(t) =g_i\left(z_{i 2}(t)+\alpha_{i 1}(t)\right)+F_{i 1}\left(X_{i 1}(t)\right)+D_{i 1}(t) \\
\nonumber & =g_i z_{i 2}(t)+g_i \hat{v}_{i 1}^*(t) +g_i h_{i 1}(t)+F_{i 1}\left(X_{i 1}(t)\right)+D_{i 1}(t) \\
\nonumber & =g_i z_{i 2}(t)-k_{i 1} e_i(t)-\hat{W}_{a, i 1}^\top(t) S_{i 1}\left(e_i(t)\right)+g_i h_{i 1}(t) \\
&+F_{i 1}\left(X_{i 1}(t)\right) +D_{i 1}(t).
\end{align}

Let $F_{i 1}\left(X_{i 1}(t)\right)$ be represented as,
\begin{equation}
\begin{aligned}
    F_{i 1}\left(X_{i 1}(t)\right)&={\theta_{i 1}^*}^\top \Phi_{i 1}\left(x_{i 1}(t)\right)+\delta_{i 1}\left(X_{i 1}(t)\right),
\end{aligned}  
\end{equation}
where, $\theta_{i 1}^*$ represents the vector of optimal weights, and $\Phi_{i1}$ denotes the vector of neural network activation functions associated with the approximation of unknown functions in the initial step. The function $F_{i1}$ incorporates the first and second state variables of the neighboring nodes. As the remaining state variables are also present in subsequent steps due to the derivative of the virtual control law, based on the properties of the neural network, we decrease the number of inputs from $X_{i1}$ to $x_{i1}$, ensuring that the proposed controller requires only the output variable of the neighboring nodes. \\
Consequently, the second component of the control policy, that is, $h_{i1} (t)$, responsible for ensuring fixed-time consensus and compensation of uncertainties, is proposed as,
\begin{equation}
\begin{aligned}
    h_{i 1}(t) & =\frac{1}{g_i} \left( -k_{p, i 1} e_i^p(t)-k_{q, i 1} e_i^q(t) \right. \\
    & \left. -\hat{\theta}^\top_{i 1}(t) \phi_{i 1}\left(x_{i 1}(t)\right)-\hat{D}_{i 1}(t) \right),
\end{aligned}
\end{equation}
With $k_{p,i1},k_{q,i1}>0$. Finally, the terms $\hat{\theta}_{i 1}(t)$ and $\hat{D}_{i 1}(t)$  are estimates of ${\theta}^*_{i1} (t)$ and $D_{i1}$, with adaptation laws,
\begin{equation}
    \dot{\hat{\theta}}_{i 1}(t)=\Gamma_{\theta, i 1}\left(e_i(t) \phi_{i 1}\left(x_{i 1}(t)\right)-\sigma_{1 \theta, i 1} \hat{\theta}_{i 1}^{{p}}(t)-\sigma_{2 \theta, i 1} \hat{\theta}_{i 1}^q(t)\right),
\end{equation}
 \begin{equation}
   \dot{\hat{D}}_{i 1}(t)=\gamma_{D, i 1}\left(e_i(t)-\sigma_{1 D, i 1} \hat{D}_{i 1}^p(t)-\sigma_{2 D, i 1} \hat{D}_{i 1}^q(t)\right),
 \end{equation}
where $\sigma_{1 \theta, i 1}, \sigma_{2 \theta, i 1},\sigma_{1 D, i 1},\sigma_{2 D, i 1}, \gamma_{D, i 1} >0$, and $\Gamma_{\theta, i 1} \succ 0$. Note that the consensus dynamics of the first step can now be represented as,
\begin{equation}
    \begin{aligned}
     &\dot{e}_i(t)  =-k_{i 1} e_i(t)-k_{{p}, i 1} e_i^{p}(t)-k_{q, i 1} e_i^q(t) \\
     & -\hat{W}_{a, i 1}^\top(t) S_{i 1}\left(e_i(t)\right)-{\tilde{\theta}}^\top_{i 1}(t) \Phi_{i 1}\left(x_{i 1}(t)\right) +\delta_{i 1}\left(x_{i 1}(t)\right) \\
     & -\widetilde{D}_{i 1}(t)+g_i z_{i 2}(t). 
    \end{aligned}
\end{equation}

Now consider the following Lyapunov function for Step 1, where $\tilde{W}_{c,i1} (t) =\hat{W}_{c,i1}(t)-W^*_{i1}$, $\tilde{W}_{a,i1} (t) =\hat{W}_{a,i1}(t)-W^*_{i1}$, $\tilde{\theta}_{i1} (t) =\hat{\theta}_{i1}(t)-{\theta}^*_{i1}$ and $\tilde{D}_{i1} (t) =\hat{D}_{i1}(t)-D_{i1}$ are the
estimation error of the critic and actor neural network weights, the approximation of the unknown function, and the
disturbance estimation error, respectively.

\begin{equation}\label{eq:lyap1}
\begin{aligned}
&V_{i 1}  =\frac{1}{2} e_i^2(t)+\frac{1}{2} \tilde{W}_{c, i 1}^\top(t) \Gamma_{c, i 1}^{-1} \tilde{W}_{c, i 1}(t) \\
& +\frac{1}{2} \tilde{W}_{a, i 1}^\top(t)\Gamma_{a, i 1}^{-1} \tilde{W}_{a, i 1}(t)+\frac{1}{2} \tilde{\theta}_{i 1}^\top(t) \Gamma_{\theta, i 1}^{-1} \tilde{\theta}_{i 1}(t)  + \frac{1}{2 \gamma_{D, i 1}} \tilde{D}_{i 1}^2(t).
\end{aligned}
\end{equation}

The time derivative of the presented Lyapunov function is given by, 
\begin{equation} \label{eq:lyapder}
    \begin{aligned}
&\dot{V}_{i 1}  =e_i(t) \dot{e}_i(t) +\tilde{W}^\top_{c, i 1}(t) \Gamma_{c, i 1}^{-1} \dot{\hat{W}}_{c, i 1}(t) \\
 & +\tilde{W}^\top_{a, i 1}(t) \Gamma_{a, i 1}^{-1} \dot{\hat{W}}_{a, i 1}(t) +\tilde{\theta}^\top_{i 1}(t) \Gamma_{\theta, i 1}^{-1} \dot{\hat{\theta}}_{i 1}(t) \\
& +\frac{1}{\gamma_{D, i 1}} \tilde{D}_{i 1}(t) \dot{\hat{D}}_{i 1}(t)-\frac{1}{\gamma_{D, i 1}} \tilde{D}_{i 1}(t) \dot{D}_{i 1}(t).
\end{aligned}
\end{equation}
Lemma~\ref{lemma:Lyap1} establishes an upper bound on \eqref{eq:lyapder}, which is subsequently used in the fixed-time consensus analysis of the multi-agent system.

\begin{lemma}\label{lemma:Lyap1}
    Let the control and adaptation parameters  of the first step meet the following conditions,
\begin{equation}
\begin{aligned}
    & k_{i1} > \tfrac{3}{2}, \quad \sigma_{1c,i1} > \sigma_{1a,i1} > 1, \quad \sigma_{2c,i1} > 2\sigma_{2a,i1} > 0, \\
    & 0 < \sigma_{3a,i1} < \tfrac{\sigma_{3c,i1}}{8 \times 172}, \quad  \sigma_{2D,i1} > \tfrac{2}{\mu_{D,i1}^4}.
\end{aligned}
\label{eq:lemma9_parameters}
\end{equation}
    Then the time derivative of the Lyapunov function \eqref{eq:lyapder} meets the inequality \eqref{eq:lyap1inequality},

\begin{equation}\label{eq:lyap1inequality}
\begin{aligned}
\dot{V}_{i1} \leq\ & 
- k_{p,i1} e_i^{p+1} - k_{q,i1} e_i^{q+1} \\
& - \left( \sigma_{2ca,i1} \left\| \tilde{W}_{c,i1}^{\left(\frac{p+1}{2}\right)} \right\|^2 + \sigma_{3ca,i1} \left\| \tilde{W}_{c,i1}^{\left(\frac{q+1}{2}\right)} \right\|^2 \right) \\
& - \left( \tfrac{1}{2} \sigma_{2a,i1} \left\| \tilde{W}_{a,i1}^{\left(\frac{p+1}{2}\right)} \right\|^2 + \underline{\sigma}_{3a,i1} \left\| \tilde{W}_{a,i1}^{\left(\frac{q+1}{2}\right)} \right\|^2 \right) \\
& - \left( \tfrac{1}{2} \sigma_{1\theta,i1} \left\| \tilde{\theta}_{i1}^{\left(\frac{p+1}{2}\right)} \right\|^2 + \underline{\sigma}_{2\theta,i1} \left\| \tilde{\theta}_{i1}^{\left(\frac{q+1}{2}\right)} \right\|^2 \right) \\
& - \left( \tfrac{1}{2} \sigma_{1D,i1} \tilde{D}_{i1}^{p+1} + \tilde{\sigma}_{2D,i1} \tilde{D}_{i1}^{q+1} \right) + C_{i1} + g_i z_{i2} e_i,
\end{aligned}
\end{equation}

where, 
\begin{equation*}
\begin{aligned}
    & \sigma_{2ca,i1} = \tfrac{1}{2} \sigma_{2c,i1} - \sigma_{2a,i1}, \quad 
    \sigma_{3ca,i1} = \underline{\sigma}_{3c,i1} - \bar{\sigma}_{3a,i1}, \\
    & \underline{\sigma}_{3c,i1} = \tfrac{1}{8} \sigma_{3c,i1}, \quad 
    \bar{\sigma}_{3a,i1} = 172\, \sigma_{3a,i1}, \quad 
    \underline{\sigma}_{3a,i1} = \tfrac{1}{8} \sigma_{3a,i1}, \\
    & \underline{\sigma}_{2\theta,i1} = \tfrac{1}{8} \sigma_{2\theta,i1}, \quad 
    \tilde{\sigma}_{2D,i1} = \underline{\sigma}_{2D,i1} - \tfrac{1}{4\mu_{D,i1}^4}, \quad 
    \underline{\sigma}_{2D,i1} = \tfrac{1}{8} \sigma_{2D,i1}.
\end{aligned}
\end{equation*}

Moreover, $C_{i1}$ encompasses constant terms, i.e., the weights of the neural networks and the bounds of external disturbances, and is given as,
\begin{equation*}
\begin{aligned}
C_{i 1} =\; & \frac{1}{2} \bar{\delta}_{i 1} 
+ \frac{\sigma_{1 c, i 1}}{2} \, \bar{\lambda}_{s, i 1} \left\| W_{i 1}^* \right\|^2 + \sigma_{2 c, i 1} \left\| {W_{i 1}^*}^{\frac{p+1}{2}} \right\|^2  \\
& 
+ \bar{\sigma}_{3 c, i 1} \left\| {W_{i 1}^*}^{\frac{q+1}{2}} \right\|^2  + \sigma_{1 \theta, i 1} \left\| {\theta_{i 1}^*}^{\frac{p+1}{2}} \right\|^2 
+ \bar{\sigma}_{2 \theta, i 1} \left\| {\theta_{i 1}^*}^{\frac{q+1}{2}} \right\|^2 \\
& + \sigma_{1 D, i 1} \, \bar{D}_{i 1}^{p+1} 
+ \bar{\sigma}_{2 D, i 1} \, \bar{D}_{i 1}^{q+1} 
+ \frac{3}{4} \left(\frac{\mu_{D, i 1}}{\gamma_{D, i 1}}\right)^{\frac{4}{3}} \bar{D}_{d, i 1}^{\frac{4}{3}}.
\end{aligned}
\end{equation*}

The term $\bar{\lambda}_{s, i 1}$ corresponds the largest eigenvalue of $S_{i1}(z_{i1}(t))S^\top_{i1}(z_{i1}(t))$. Moreover, 
\begin{equation*}
\begin{aligned}
    & \bar{\sigma}_{3c,i1} = 172\, \sigma_{3c,i1}, \quad 
    \bar{\sigma}_{2\theta,i1} = 172\, \sigma_{2\theta,i1}, \quad 
    \bar{\sigma}_{2D,i1} = 172\, \sigma_{2D,i1}.
\end{aligned}
\end{equation*}
The term $\mu_{D,i1}$ denotes a positive constant. The quantities $\bar{D}_{i1}$ and $\bar{D}_{d,i1}$ are the bounds of $D_{i1}(t)$ and $\dot{D}_{i1}(t)$, respectively, such that $|D_{i1}(t)| < \bar{D}_{i1}$ and $|\dot{D}_{i1}(t)| < \bar{D}_{d,i1}$. These bounds are specified in Assumption 3. 
\end{lemma}
\noindent\textbf{Proof.} The proof is given in Appendix C.

\subsection{Step $j=2,\dots,n$ for RL-based backstepping}
In this section, the developed virtual policies for steps $j=2,\ldots n-1$, and the actual control policy for step $n$ are derived. To this end, note that the dynamics of the virtual error of the agent $j$, at step $j$, i.e., $z_{ij}$, and the corresponding virtual control ${\alpha}_{i(j-1)}$ can be represented as,
\begin{align*}
    & \dot{z}_{i j}(t)=x_{i(j+1)}(t)+f_{i j}\left(\bar{x}_{i j}(t)\right)+d_{i j}(t)-\dot{\alpha}_{i(j-1)}(t) \\
    & \dot{\alpha}_{i(j-1)}(t)=\dot{\hat{v}}_{i(j-1)}^*(t)+\dot{h}_{i(j-1)}(t)=F_{\alpha, i(j-1)}\left(X_{i j}(t)\right),
\end{align*}
where $X_{ij}(t)$ represents the concatenated state vector in step $j$, given as,
\begin{align*}
    & X_{ij}(t) = \left[\bar{x}_{i j}^\top(t),~ \bar{x}_{l 2}^\top(t),~ \bar{x}_{02}^\top(t),~ \hat{W}_{c, i 1}^\top(t)~ \cdots, ~\hat{W}_{c, i(j-1)}^\top(t),\right.\\
    &\quad\hat{W}_{a, i 1}^\top(t)~ \cdots, \hat{W}_{a, i(j-1)}^\top(t),~\hat{\theta}_{i 1}^\top(t)~ \cdots,~ \hat{\theta}_{i(j-1)}^\top(t), \\ 
    &\quad\left.\hat{D}_{i 1}(t), \hat{D}_{i 2}(t), \cdots, \hat{D}_{i(j-1)}(t)\right]^\top. 
\end{align*}

In this setting, ${\bar{x}}_{l2}$ represents the first two state variables of $l$th neighbor of follower $i$. Similarly, ${\bar{x}}_{02}$ is applied only for the followers directly connected to the leader. Accordingly, the virtual error dynamics are given as,
\begin{align*}
    &\dot{z}_{i j}(t)= x_{i(j+1)}(t)+F_{i j}\left(X_{i j}(t)\right)+d_{i j}(t),\\
    &F_{ij}(X_{ij}(t))= f_{ij}({\bar{x}}_{ij}(t))-F_{\alpha,i(j-1)}(X_{ij}(t)).
\end{align*}

Similar to step 1, the virtual control policy of the $j$th step $\alpha_{ij}(t)$ includes $v^*_{ij} (t)$ and $h_{ij}(t)$, where $v^*_{ij} (t)$  is derived in an optimal manner and $h_{ij}(t)$ guarantees the fixed-time consensus of the closed-loop system, that is,
\begin{equation}\label{eq:virtualj}
    \alpha_{ij}(t)=v^*_{ij} (t)+h_{ij}(t).
\end{equation}
For the last step, the actual control policy corresponds to $\alpha_{in}(t)$, that is,
$u_{i}(t)=\alpha_{in}(t)$. The performance index of the $j$th step is defined as,
\begin{equation}
    J_{i j}\left(z_{i j}(t)\right)=\int_t^{\infty}\left(z_{i j}^2(s)+v_{i j}^2(s)\right) d s.
\end{equation}
The HJB equation in step $j$ is derived as,
\begin{equation}
\begin{aligned}
& H_{i j}\left(z_{i j}(t), v_{i j}^*(t), J_{i j}^*\right)  =z_{i j}^2(t)+v_{i j}^{* 2}(t) \\
&+\frac{d J_{i j}^*\left(z_{i j}(t)\right)}{d z_{i j}(t)}\left(v_{i j}^*(t)+F_{i j}\left(X_{i j}(t)\right)+d_{i j}(t)\right)=0.
\end{aligned}   
\end{equation}
Similarly, the optimal control law $v^*_{ij} (t)$ is given by $ v_{i j}^*(t)=-\frac{1}{2} \frac{d J_{i j}^*\left(z_{i j}(t)\right)}{d z_{i j}(t)}$. After decomposing $\frac{d J_{i j}^*\left(z_{i j}(t)\right)}{d z_{i j}(t)}$ into the form,
\begin{equation}
    \frac{d J_{i j}^*\left(z_{i j}(t)\right)}{d z_{i j}(t)}=2 k_{i j} z_{i j}(t)+2 J_{i j}^r\left(z_{i j}(t)\right),
\end{equation}
the optimal control policy $v_{i j}^*(t)$ can be written as,
\begin{equation}
    v_{i j}^*(t)=-k_{i j} z_{i j}(t)-J_{i j}^r\left(z_{i j}(t)\right).
\end{equation}
Next, an RL-based neural network is utilized to approximate the unknown term $J_{i j}^r\left(z_{i j}(t)\right)$,
\begin{equation}
    J_{i j}^r\left(z_{i j}(t)\right)={W_{i j}^{*}}^\top S_{i j}\left(z_{i j}(t)\right)+\varepsilon_{i j}\left(z_{i j}(t)\right),
\end{equation}
with the optimal weights $W_{i j}^{*}$ and the estimation error $\varepsilon_{i j}$. Accordingly, 
approximations of the optimal values $\frac{d J_{i j}^*\left(z_{i j}(t)\right)}{d z_{i j}(t)}$ and $v_{i j}^*(t)$ satisfy,
\begin{equation}
    \frac{d \hat{J}_{i j}^*\left(z_{i j}(t)\right)}{d z_{i j}(t)}=2 k_{i j} z_{i j}(t)+2 {\hat{W}_{c, i j}}^\top S_{i j}\left(z_{i j}(t)\right)  
\end{equation}
\begin{equation}
   \hat{v}_{i j}^*(t)=-k_{i j} z_{i j}(t)-{\hat{W}_{a, i j}}^\top(t) S_{i j}\left(z_{i j}(t)\right),
\end{equation}
where $\hat{W}_{c, i j}$ and $\hat{W}_{a, i j}$ are the estimates of the ideal weights of the actor-critic neural network in the $j$th step. 

Consider the following update rules of step $j$ for the weights of the actor and critic neural network, respectively.
\begin{equation}
\left\{
\begin{aligned}
   & \dot{\hat{W}}_{c, i j}(t)  =\Gamma_{c, i j}\left[-S_{i j}\left(z_{i j}(t)\right) z_{i j}(t)\right. \\
  & \left. -\sigma_{1 c, i j} S_{i j}\left(z_{i j}(t)\right) S_{i j}^\top\left(z_{i j}(t)\right) \hat{W}_{c, i j}(t) \right. \\
  & \left.-\sigma_{2 c, i j} \hat{W}_{c, i j}^p(t)-\sigma_{3 c, i j} \hat{W}_{c, i j}^q(t)\right],\\
    & \dot{\hat{W}}_{a, i j}(t) =\\
    &\Gamma_{a, i j}\left[-\sigma_{1 a, i j} S_{i j}\left(z_{i j}(t)\right) S_{i j}^\top\left(z_{i j}(t)\right)\left(\hat{W}_{a, i j}(t)-\hat{W}_{c, i j}(t)\right) \right. \\
   & -\sigma_{2 a, i j}\left(\hat{W}_{a, i j}(t)-\hat{W}_{c, i j}(t)\right)^p\\
   &\left.-\sigma_{3 a, i j}\left(\hat{W}_{a, i j}(t)-\hat{W}_{c, i j}(t)\right)^q \right]. \label{eq:actor_criticj}
\end{aligned}
\right.
\end{equation}
Analogous to the analysis in Step 1, the first term in the adaptation law for the actor network is derived based on a gradient descent principle. The next stage is to devise the structure of the fixed-time stabilizer term and its associated adaptation laws in the context of the $j$th step’s virtual control law. By incorporation of the actor weights and virtual control input into the virtual error dynamics, the reformulated dynamics of ${z}_{i j}$ is derived as,
\begin{equation}
    \begin{aligned}
 & \dot{z}_{i j}(t)=\alpha_{i j}(t)+ z_{i(j+1)}(t)+F_{i j}\left(X_{i j}(t)\right)+d_{i j}(t) \\
  & =\hat{v}_{i j}^*(t)+h_{i j}(t)+F_{i j}\left(X_{i j}(t)\right)+d_{i j}(t)+z_{i(j+1)}(t) \\
  & =-k_{i j} z_{i j}(t)-\hat{{W}}_{a, i j}^\top(t) S_{i j}\left(z_{i j}(t)\right)+h_{i j}(t)+F_{i j}\left(X_{i j}(t)\right) \\
  & +d_{i j}(t)+z_{i(j+1)}(t).
\end{aligned}
\end{equation}
Note that for the step $n$, we have, $z_{i(n+1)}(t)=0$. The unknown term $F_{ij}(X_{ij}(t))$ is approximated using the neural network as,
\begin{equation}
    \begin{aligned}
&F_{i j}\left(X_{i j}(t)\right)={\theta_{F, i j}^*}^\top \Phi_{F, i j}\left(X_{i j}(t)\right) +\delta_{F, i j}\left(X_{i j}(t)\right) \\
&={\theta_{i j}^*}^\top \Phi_{i j}\left(\chi_{i j}(t)\right)+\delta_{i j}\left(X_{i j}(t)\right),
\end{aligned}
\end{equation}
where $\chi_{i j}(t)$ is the truncated input of the network approximating the unknown function of the $j$th step given by,
\begin{align*}
    &\chi_{i j}(t) =\left[\bar{x}_{i j}^\top(t),~  \hat{W}_{c, i 1}^\top(t)~ \cdots, ~\hat{W}_{c, i(j-1)}^\top(t),\right.\\
    &\quad\hat{W}_{a, i 1}^\top(t)~ \cdots, \hat{W}_{a, i(j-1)}^\top(t),~\hat{\theta}_{i 1}^\top(t)~ \cdots,~ \hat{\theta}_{i(j-1)}^\top(t), \\
    &\quad\left.\hat{D}_{i 1}(t), \hat{D}_{i 2}(t), \cdots, \hat{D}_{i(j-1)}(t)\right]^\top, 
\end{align*}
and $\delta_{ij}(X_{ij}(t))={\theta_{F, i j}^*}^\top \Phi_{F, i j}\left(X_{i j}(t)\right)-\theta_{i j}^*{ }^\top \Phi_{i j}\left(\chi_{i j}(t)\right)+\delta_{F, i j}\left(X_{i j}(t)\right)$ is the neural network estimation error with truncated inputs. The ideal weights and activation functions of the neural network with truncated inputs are $\theta^*_{ij}$ and $\Phi_{ij}$, respectively. The second component of the control policy $h_{ij} (t)$, is proposed as,
\begin{equation}
\begin{aligned}
    & h_{i j}(t)=-k_{p, i j} z_{i j}^p(t)-k_{q, i j} z_{i j}^q(t) \\
    & -\hat{\theta}_{i j}^\top(t) \Phi_{i j}\left(\chi_{i j}(t)\right)-\hat{d}_{i j}(t)-z_{i(j-1)}^r(t),
\end{aligned}
\end{equation}
with,
\begin{equation}
    z^r_{i1} (t)=g_{i} e_{i}(t),~~ z^r_{i (j-1)} (t)=z_{i (j-1)}(t), \qquad  \forall j \in \{3, \ldots n\},
\end{equation}
and the adaptation laws are proposed as,
\begin{equation}
 \dot{\hat{\theta}}_{i j}(t)=\Gamma_{\theta, i j}\left(z_{i j}(t) \Phi_{i j}\left(\chi_{i j}(t)\right)-\sigma_{1 \theta, i j} \hat{\theta}_{i j}^p(t)-\sigma_{2 \theta, i j} \hat{\theta}_{i j}^q(t)\right),
\end{equation}
\begin{equation}
  \dot{\hat{d}}_{i j}(t)=\gamma_{d, i j}\left(z_{i j}(t)-\sigma_{1 d, i j} {\hat{d}}_{i j}^p(t)-\sigma_{2 d, i j} {\hat{d}}_{i j}^q(t)\right),
\end{equation}
with positive parameters $k_{ij}$, $\sigma_{1c,ij}$, $\sigma_{2c,ij}$, $\sigma_{3c,ij}$, $\sigma_{1a,ij}$, $\sigma_{2a,ij}$, $\sigma_{3a,ij}$, $\sigma_{1\theta,ij}$, $\sigma_{2\theta,ij}$, $\sigma_{1d,ij}$, $\sigma_{2d,ij}$, and $\gamma_{d,ij}$ and positive-definite matrices $\Gamma_{c,ij}$, $\Gamma_{a,ij}$ and $\Gamma_{\theta,ij}$.\\
Incorporating the derived policies in the virtual error dynamics gives,
\begin{equation}
    \begin{aligned}
       & \dot{z}_{i j}(t)=-k_{i j} z_{i j}(t)-k_{p, i j} z_{i j}^p(t)-k_{q, i j} z_{i j}^q(t) \\
       &-\hat{W}_{a, i j}^\top(t) S_{i j}\left(z_{i j}(t)\right)-z_{i(j-1)}^r(t) 
       -\tilde{\theta}_{i j}^\top(t) \Phi_{i j}\left(X_{i j}(t)\right) \\
       &-\tilde{d}_{i j}(t)+\delta_{i j}\left(X_{i j}(t)\right)+z_{i(j+1)}(t),
    \end{aligned}
\end{equation}
where $\tilde{\theta}_{ij}(t)=\hat{\theta}_{ij}(t)-{\theta}^*_{ij}$ and $\tilde{d}_{ij}(t)=\hat{d}_{ij}(t)-d_{ij}(t)$ are the estimation error of the ideal weights of the neural network and disturbance, respectively. 
Now, consider the following Lyapunov function for step $j$th, given by,
\begin{equation}\label{eq:Lyapj}
    \begin{aligned}
        & V_{i j}=V_{i(j-1)}+\frac{1}{2} z_{i j}^2(t)+\frac{1}{2} \tilde{W}_{c, i j}^\top(t) \Gamma_{c, i j}^{-1} \tilde{W}_{c, i j}(t) \\
        &+\frac{1}{2} \tilde{W}_{a, i j}^\top(t) \Gamma_{a, i j}^{-1} \tilde{W}_{a, i j}(t)+\frac{1}{2} \tilde{\theta}_{i j}^\top(t) \Gamma_{\theta, i j}^{-1} \tilde{\theta}_{i j}(t) +\frac{1}{2 \gamma_{d, i j}} \tilde{d}_{i j}^2(t),
    \end{aligned}
\end{equation}
with $\tilde{W}_{c, i j}(t)=\hat{W}_{c, i j}(t)-W^*_{ij}, \quad \tilde{W}_{a, i j}(t)=\hat{W}_{a, i j}(t)-W^*_{ij}$.
The time derivative of $V_{i j}$ is then given by,
\begin{equation}
    \begin{aligned}
        &\dot{V}_{i j}=\dot{V}_{i(j-1)}+z_{i j}(t) \dot{z}_{i j}(t)+\tilde{W}_{c, i j}^\top(t) \Gamma_{c, i j}^{-1} \dot{\hat{W}}_{c, i j}(t) \\
        & +\tilde{W}_{a, i j}^\top(t) \Gamma_{a, i j}^{-1}\dot{\hat{W}}_{a, i j}(t)+\tilde{\theta}_{i j}^\top(t) \Gamma_{\theta, i j}^{-1} \dot{\hat{\theta}}_{i j}(t) \\
        & +\frac{1}{\gamma_{d, i j}} \tilde{d}_{i j}(t) \dot{\hat{d}}_{i j}(t)-\frac{1}{\gamma_{d, i j}} \tilde{d}_{i j}(t) \dot{d}_{i j}(t).
    \end{aligned}
\end{equation}
Lemma \ref{lemma:Lyapj} gives an upper bound on $\dot{V}_{i j}$, which is subsequently used in fixed-time consensus analysis. 

\begin{lemma}\label{lemma:Lyapj}
    Let the control and adaptation parameters  of the $j$th step meet the following conditions,
\begin{equation}
\begin{aligned}
    & k_{ij} > \tfrac{3}{2}, \quad \sigma_{1c,ij} > \sigma_{1a,ij} > 1, \quad \sigma_{2c,ij} > 2\sigma_{2a,ij} > 0, \\
    & 0 < \sigma_{3a,ij} < \tfrac{\sigma_{3c,ij}}{8 \times 172}, \quad \sigma_{2D,ij} > \tfrac{2}{\mu_{D,ij}^4}.
\end{aligned}
\label{eq:lemma10_parameters}
\end{equation}
Then the time derivative of the Lyapunov function \eqref{eq:Lyapj} meets the inequality \eqref{eq:lyapjinequality},

\begin{equation}\label{eq:lyapjinequality}
\begin{aligned}
\dot{V}_{ij} \leq\ & 
- \sum_{k=1}^j \left( k_{p,ik} z_{ik}^{r(p+1)} + k_{q,ik} z_{ik}^{r(q+1)} \right) \\
& - \sum_{k=1}^j \left( \sigma_{2ca,ik} \left\| \tilde{W}_{c,ik}^{\left(\frac{p+1}{2}\right)} \right\|^2 + \sigma_{3ca,ik} \left\| \tilde{W}_{c,ik}^{\left(\frac{q+1}{2}\right)} \right\|^2 \right) \\
& - \sum_{k=1}^j \left( \tfrac{1}{2} \sigma_{2a,ik} \left\| \tilde{W}_{a,ik}^{\left(\frac{p+1}{2}\right)} \right\|^2 + \sigma_{3a,ik} \left\| \tilde{W}_{a,ik}^{\left(\frac{q+1}{2}\right)} \right\|^2 \right) \\
& - \sum_{k=1}^j \left( \tfrac{1}{2} \sigma_{1\theta,ik} \left\| \tilde{\theta}_{ik}^{\left(\frac{p+1}{2}\right)} \right\|^2 + \underline{\sigma}_{2\theta,ik} \left\| \tilde{\theta}_{ik}^{\left(\frac{q+1}{2}\right)} \right\|^2 \right) \\
& - \sum_{k=1}^j \left( \tfrac{1}{2} \sigma_{1d,ik}^r \tilde{d}_{ik}^{r(p+1)} + \sigma_{2d,ik}^r \tilde{d}_{ik}^{r(q+1)} \right) \\
& + \sum_{k=1}^j C_{ik} + z_{i(j+1)} z_{ij},
\end{aligned}
\end{equation}

where,

\begin{equation}
\begin{aligned}
    & \sigma_{2ca,ij} = \tfrac{\sigma_{2c,ij}}{2} - \sigma_{2a,ij}, \quad 
    \sigma_{3ca,ij} = \underline{\sigma}_{3c,ij} - \overline{\sigma}_{3a,ij}, \quad
    \underline{\sigma}_{3c,ij} = \tfrac{\sigma_{3c,ij}}{8}, \\
    & \overline{\sigma}_{3a,ij} = 172\, \sigma_{3a,ij}, \quad 
    \underline{\sigma}_{3a,ij} = \tfrac{\sigma_{3a,ij}}{8}, \quad
    \underline{\sigma}_{2\theta,ij} = \tfrac{\sigma_{2\theta,ij}}{8}, \\
    & \tilde{\sigma}_{2D,ij} = \underline{\sigma}_{2D,ij} - \tfrac{1}{4 \mu_{D,ij}^4}, \quad 
    \underline{\sigma}_{2D,ij} = \tfrac{\sigma_{2D,ij}}{8}.
\end{aligned}
\end{equation}
The term $C_{ij}$ contains constant terms derived from the unknown weights of neural networks and the unknown bound of external disturbances and is defined as,

\begin{equation*}
\begin{aligned}
C_{ij} =\ &
\tfrac{1}{2} \sigma_{1c,ij} \bar{\lambda}_{s,ij} \left\| W_{ij}^* \right\|^2 
+ \sigma_{2c,ij} \left\| W_{ij}^{*\left(\frac{p+1}{2}\right)} \right\|^2 
+ \bar{\sigma}_{3c,ij} \left\| W_{ij}^{*\left(\frac{q+1}{2}\right)} \right\|^2 \\
& + \sigma_{1\theta,ij} \left\| \theta_{ij}^{*\left(\frac{p+1}{2}\right)} \right\|^2 
+ \bar{\sigma}_{2\theta,ij} \left\| \theta_{ij}^{*\left(\frac{q+1}{2}\right)} \right\|^2 \\
& + \sigma_{1d,ij} \bar{d}_{ij}^{p+1} 
+ \sigma_{2d,ij} \bar{d}_{ij}^{q+1} 
+ \tfrac{3}{4} \left( \tfrac{\mu_{d,ij}}{\gamma_{d,ij}} \right)^{\frac{4}{3}} \bar{d}_{d,ij}^{4/3} 
+ \tfrac{1}{2} \bar{\delta}_{ij}.
\end{aligned}
\end{equation*}

where $\bar{\lambda}_{s, i j}$ corresponds to the largest eigenvalue of $S_{ij} (z_{ij}(t)) S_{ij}^\top (z_{ij}(t))$. Moreover,
\begin{equation}
\begin{aligned}
    & \bar{\sigma}_{3c,ij} = 172\, \sigma_{3c,ij}, \quad
    \bar{\sigma}_{2\theta,ij} = 172\, \sigma_{2\theta,ij}, \quad
    \bar{\sigma}_{2D,ij} = 172\, \sigma_{2D,ij}.
\end{aligned}
\end{equation}

Moreover, $\mu_{D, i j}$ is a positive constant and
\begin{equation*}
    z_{i k}^r(t)=\left\{\begin{array}{l}g_i e_i(t), ~~ k=1 \\ z_{i k}(t), ~ 2\leq k \leq j\end{array},\right.
    \sigma_{1 d, i k}^r= \left\{\begin{array}{l}\sigma_{1 D, i 1}, ~ k=1 \\ \sigma_{1 d, i k}, ~ 2\leq k \leq j\end{array},\right.
\end{equation*}

\begin{equation*}
    \bar{\sigma}_{2d, ik}^r(t)=\left\{\begin{array}{l}\bar{\sigma}_{2 D, i j}, ~~ k=1 \\ \bar{\sigma}_{2 d, i j},  2\leq k \leq j\end{array},\right.
  \tilde{d}^r_{ik}(t)= \left\{\begin{array}{l}\tilde{D}_{i1}(t),  k=1 \\ \tilde{d}_{ik}(t),  2\leq k \leq j\end{array}.\right.
\end{equation*}
\end{lemma}
\noindent\textbf{Proof.} The proof is given in Appendix D.

\section{Fixed-time Consensus Analysis}
Assume that each agent incorporates a reinforcement learning algorithm based on the adaptation laws \eqref{eq:actor_critic1}, \eqref{eq:actor_criticj}, constructs the virtual control policies according to \eqref{eq:virtual1}, \eqref{eq:virtualj}, and applies the actual control policy $u_i=\alpha_{in}$. The following theorem shows that the proposed control algorithm ensures fixed-time consensus. In this case, the followers would follow the leader in a fixed time with a bounded consensus error.

\begin{theorem}
Consider applying the proposed RL-backstepping fixed-time control strategy  to the leader-follower dynamics \eqref{eq:leader}, \eqref{eq:followers}. Then the followers are guaranteed to track the leader in $T<T_{\max } \leq \frac{1}{k_p(1-p)}+\frac{1}{k_q(q-1)}$ independent of the initial conditions in the presence of uncertainties and disturbances so that the tracking error converges to a neighborhood of the origin with adjustable convergence radius $\Omega_z$.
\end{theorem}

\textbf{Proof:} Consider the overall Lyapunov function,

\begin{equation}
    \begin{aligned}
& V =\sum_{i=1}^N\left( \frac{1}{2} e_i^2(t)+\sum_{k=2}^n \frac{1}{2} z_{i k}^2(t) +\sum_{k=1}^n\left(\frac{1}{2} \tilde{W}_{c, i k}^\top(t) \Gamma_{c, i k}^{-1} \tilde{W}_{c, i k}(t)\right.\right. \\
& + \left. \left.\frac{1}{2} \tilde{W}_{a, i k}^\top(t) \Gamma_{a, i k}^{-1} \tilde{W}_{a, i k}(t)+\frac{1}{2} \tilde{\theta}_{i k}^\top(t) \Gamma_{\theta, i k}^{-1} \tilde{\theta}_{i k}(t)+\frac{1}{2 \gamma_{d, i k}} \tilde{d}_{i k}^2(t)\right)\right)
\end{aligned}
\end{equation}

According to Lemma \ref{lemma:Lyap1} and \ref{lemma:Lyapj}, the time derivative of $V$ satisfies the following inequality,

\begin{equation*}
\begin{aligned}
&\dot{V} \leq -\sum_{i=1}^N \sum_{k=1}^n \Bigg(
 k_{p,ik} z_{ik}^{r(p+1)} + k_{q,ik} z_{ik}^{r(q+1)} \\
& + \sigma_{2ca,ik} \left\| \tilde{W}_{c,ik}^{\left(\frac{p+1}{2}\right)} \right\|^2
+ \sigma_{3ca,ik} \left\| \tilde{W}_{c,ik}^{\left(\frac{q+1}{2}\right)} \right\|^2 + \tfrac{1}{2} \sigma_{2a,ik} \left\| \tilde{W}_{a,ik}^{\left(\frac{p+1}{2}\right)} \right\|^2
\end{aligned}
\end{equation*}

\begin{equation}
\begin{aligned}
&+ \sigma_{3a,ik} \left\| \tilde{W}_{a,ik}^{\left(\frac{q+1}{2}\right)} \right\|^2 + \tfrac{1}{2} \sigma_{1\theta,ik} \left\| \tilde{\theta}_{ik}^{\left(\frac{p+1}{2}\right)} \right\|^2
+ \sigma_{2\theta,ik} \left\| \tilde{\theta}_{ik}^{\left(\frac{q+1}{2}\right)} \right\|^2\\
& + \tfrac{1}{2} \sigma_{1d,ik}^r \tilde{d}_{ik}^{r(p+1)}
+ \sigma_{2d,ik}^r \tilde{d}_{ik}^{r(q+1)} + C_{ik} \Bigg).
\end{aligned}
\end{equation}

Let $V_1, V_2, V_3, V_4, V_5$ be defined as,
\begin{equation*}
    \begin{aligned} &V_1=\sum_{i=1}^N\left(\frac{1}{2} e_i^2(t)+\sum_{k=2}^n \frac{1}{2} z_{i k}^2(t)\right), \\
    & V_2=\sum_{i=1}^N \sum_{k=1}^n \frac{1}{2} \tilde{W}_{c, i k}^\top(t) \Gamma_{c, i k}^{-1} \tilde{W}_{c, i k}(t), \\ & V_3=\sum_{i=1}^N \sum_{k=1}^n \frac{1}{2} \tilde{W}_{a, i k}^\top(t) \Gamma_{a, i k}^{-1} \tilde{W}_{a, i k}(t), \\ & V_4=\sum_{i=1}^N \sum_{k=1}^n \frac{1}{2} \tilde{\theta}_{i k}^\top(t) \Gamma_{\theta, i k}^{-1} \tilde{\theta}_{i k}(t), \\ & V_5=\sum_{i=1}^N \sum_{k=1}^n \frac{1}{2 \gamma_{d, i k}} \tilde{d}_{i k}^2(t).
  \end{aligned}
\end{equation*}

According to Lemma 2,

\begin{equation}
\begin{aligned}
& -\sum_{i=1}^N \sum_{k=1}^n \left( k_{p,ik} z_{ik}^{r(p+1)} + k_{q,ik} z_{ik}^{r(q+1)} \right) \leq -k_p V_1^{\frac{p+1}{2}} - k_q V_1^{\frac{q+1}{2}}, \\
& -\sum_{i=1}^N \sum_{k=1}^n \left( \sigma_{2ca,ik} \left\| \tilde{W}_{c,ik}^{\left(\frac{p+1}{2}\right)} \right\|^2 + \sigma_{3ca,ik} \left\| \tilde{W}_{c,ik}^{\left(\frac{q+1}{2}\right)} \right\|^2 \right)\\
&\leq -\sigma_{2ca} V_2^{\frac{p+1}{2}} - \sigma_{3ca} V_2^{\frac{q+1}{2}}, \\
& -\sum_{i=1}^N \sum_{k=1}^n \left( \tfrac{1}{2} \sigma_{2a,ik} \left\| \tilde{W}_{a,ik}^{\left(\frac{p+1}{2}\right)} \right\|^2 + \sigma_{3a,ik} \left\| \tilde{W}_{a,ik}^{\left(\frac{q+1}{2}\right)} \right\|^2 \right)\\
&\leq -\sigma_{2a} V_3^{\frac{p+1}{2}} - \sigma_{3a} V_3^{\frac{q+1}{2}}, \\
& -\sum_{i=1}^N \sum_{k=1}^n \left( \tfrac{1}{2} \sigma_{1\theta,ik} \left\| \tilde{\theta}_{ik}^{\left(\frac{p+1}{2}\right)} \right\|^2 + \sigma_{2\theta,ik} \left\| \tilde{\theta}_{ik}^{\left(\frac{q+1}{2}\right)} \right\|^2 \right)\\
&\leq -\sigma_{1\theta} V_4^{\frac{p+1}{2}} - \sigma_{2\theta} V_4^{\frac{q+1}{2}}, \\
& -\sum_{i=1}^N \sum_{k=1}^n \left(\tfrac{1}{2} \sigma_{1d,ik}^r \tilde{d}_{ik}^{r(p+1)} + \sigma_{2d,ik}^r \tilde{d}_{ik}^{r(q+1)} \right) \leq -\sigma_{1d}^r V_5^{\frac{p+1}{2}} - \sigma_{2d}^r V_5^{\frac{q+1}{2}},
\end{aligned}
\end{equation}

$$\begin{aligned} & k_p=2^{\frac{(p+1)}{2}} \underset{i,k}{\min} \left\{k_{p, i k}\right\},  k_q=2^{\frac{(q+1)}{2}}(n N)^{1-\frac{(q+1)}{2}} \underset{i,k}{\min} \left\{k_{q, i k}\right\}, \\
& \sigma_{2 c a}=\left(\frac{2}{\bar{\lambda}_{\Gamma_c}}\right)^{\frac{(p+1)}{2}}  \underset{i,k}{\min}\left\{\sigma_{2 c a, i k}\right\},\\
&  \sigma_{3 c a}=  \left(\frac{2}{\bar{\lambda}_{\Gamma_c}}\right)^{\frac{(q+1)}{2}}(n N)^{1-\frac{(q+1)}{2}} \underset{i,k}{\min}\left\{\sigma_{3 c a, i k}\right\},\\
& \sigma_{2 a}=\frac{1}{2}\left(\frac{2}{\bar{\lambda}_{\Gamma_a}}\right)^{\frac{(p+1)}{2}} \underset{i,k}{\min}\left\{\sigma_{2 a, i k}\right\},\\
& \sigma_{3 a}=\left(\frac{2}{\bar{\lambda}_{\Gamma_a}}\right)^{\frac{(q+1)}{2}}(n N)^{1-\frac{(q+1)}{2}} \underset{i,k}{\min}\left\{\sigma_{3 a, i k}\right\},\\
& \sigma_{1 \theta}=\left(\frac{2}{\bar{\lambda}_{\Gamma_\theta}}\right)^{\frac{(p+1)}{2}} \underset{i,k}{\min}\left\{\sigma_{1 \theta, i k}\right\},\\
& \sigma_{2 \theta}=\left(\frac{2}{\bar{\lambda}_{\Gamma_\theta}}\right)^{\frac{(q+1)}{2}}(n N)^{1-\frac{(q+1)}{2}} 
\underset{i,k}{\min}\left\{\sigma_{2 \theta, i k}^1\right\},
\end{aligned}$$

$$\begin{aligned} 
& \sigma_{1 d}^r=\frac{1}{2}\left(2 \bar{\gamma}_d\right)^{\frac{(p+1)}{2}} 
\underset{i,k}{\min}\left\{\sigma_{1 d, i k}^r\right\}, \quad \bar{\gamma}_d=
\underset{i,k}{\max}\left\{\gamma_{d, i k}\right\}, \\
&\sigma_{2 d}^r=\left(2 \bar{\gamma}_d\right)^{\frac{(q+1)}{2}} 
\underset{i,k}{\min}\left\{\sigma_{2 d, i k}^r\right\}, \quad \bar{\lambda}_{\Gamma_c}=\underset{i,k}{\max}\left\{\bar{\lambda}_{\Gamma_{c, i k}}\right\},\\ &\bar{\lambda}_{\Gamma_a}=\underset{i,k}{\max}\left\{\bar{\lambda}_{\Gamma_{a, i k}}\right\}, \quad\quad\quad\quad\quad
\bar{\lambda}_{\Gamma_\theta}=\underset{i,k}{\max}\left\{\bar{\lambda}_{\Gamma_{\theta, i k}}\right\},
\end{aligned}$$

and $\bar{\lambda}_{\Gamma_{c, i k}}, \bar{\lambda}_{\Gamma_{a, i k}}, \bar{\lambda}_{\Gamma_{\theta, i k}}$ represent the maximum eigenvalues of $\Gamma_{c, i k}^{-1}, \Gamma_{a, i k}^{-1}, \Gamma_{\theta, i k}^{-1}$, respectively. Letting $\sum_{i=1}^{N} \sum_{k=1}^{n} C_{ik} = C$, the following differential inequality is obtained,

\begin{small}
    \begin{equation}
    \begin{aligned}
& \dot{V} \leq-k_p V_1^{\frac{(p+1)}{2}}-k_q V_1^{\frac{(q+1)}{2}}-\sigma_{2 c a} V_2^{\frac{(p+1)}{2}}-\sigma_{3 c a} V_2^{\frac{(q+1)}{2}} \\
& -\sigma_{2 a} V_3^{\frac{(p+1)}{2}}-\sigma_{3 a} V_3^{\frac{(q+1)}{2}}-\sigma_{1 \theta} V_4^{\frac{(p+1)}{2}}-\sigma_{2 \theta} V_4^{\frac{(q+1)}{2}} \\
& -\sigma_{1 d}^r V_5^{\frac{(p+1)}{2}} -\sigma_{2 d}^r V_5^{\frac{(q+1)}{2}}+C  \leq-\tilde k_p V^{\tilde p}-\tilde k_q V^{\tilde q}+C,
\end{aligned}
\end{equation}
\end{small}
where, $\tilde{k}_p = \min \{ k_p, \sigma_{2ca}, \sigma_{2a}, \sigma_{1\theta}, \sigma^r_{1d} \}$, $\tilde{p}=\frac{(p+1)}{2}$, $\tilde{k}_q = \min \{ k_q, \sigma_{3ca}, \sigma_{3a}, \sigma_{2\theta}, \sigma^r_{2d} \}$, $\tilde q=\frac{(q+1)}{2}$. According to Lemma 1, $V$ is attracted to the neighborhood of the origin denoted as $\Omega$ in \eqref{eq:Omega} within a fixed time $T<T_{\max } \leq \frac{1}{k_p(1-p)}+\frac{1}{k_q(q-1)}$. Note that appropriate selection of design parameters $\Gamma_{c,ik}$, $\Gamma_{a,ik}$, $\Gamma_{\theta,ik}$ and $\gamma_{d,ik}$ adjusts $\tilde k_p$ and $\tilde k_q$, and accordingly, the ultimate bound of consensus error. The consensus error is bounded within a neighborhood of the origin with radius equal to,
\begin{equation}
    \Omega_e=\sqrt{2 \min \left\{\left(\frac{C}{(1-\vartheta) k_p}\right)^p,\left(\frac{C}{(1-\vartheta) k_q}\right)^q\right\}}.
\end{equation}
Finally, the tracking error $z_{i1}$  converges within the fixed-time  to a neighborhood of the origin with the radius, 
\begin{equation}
    \Omega_z = \frac{\sqrt{2 \min \left\{\left(\frac{C}{(1-\vartheta) k_p}\right)^p,\left(\frac{C}{(1-\vartheta) k_q}\right)^q\right\}}}{\underline{\lambda}_{\tilde{L}}}. \qed
\end{equation}


\section{Simulation Case Studies} \label{sec:5}
\begin{example}\label{ex:1}
Consider the multi-agent system with the communication graph shown
in Figure \ref{fig:Topology}(a), where the leader is indicated by the index $0$. Let the dynamics of the leader and the followers be given as,
\begin{equation}
    \left\{\begin{array}{l}
    \dot{x}_{01}(t)=x_{02}(t) \\
    \dot{x}_{02}(t)=f_0\left(x_0(t)\right)+u_0(t)+d_0(t)
\end{array}\right.
\end{equation}
With $d_0 (t)= cos(t)$, $f_0(x_0(t))=50sin(x_{02}(t))$, and,
\begin{equation}
    \left\{\begin{array}{l}
    \dot{x}_{i 1}(t)=x_{i 2}(t) \\
    \dot{x}_{i 2}(t)=f_i\left(x_i(t)\right)+u_i(t)+d_i(t)
\end{array}, \quad i=1,2,3,4\right.
\end{equation}
With $d_i(t)=2sin(t)+2$ and $f_i(x_i (t))=50sin(x_{i2}(t))$,
respectively \cite{9906584}. In contrast to this study, the method in  \cite{9906584} requires the nonlinear function to be known, and the disturbance/uncertainty can be applied only to the dynamics of the last state. 

\begin{figure}[htp] 
    \centering
    \subfloat[]{%
        \includegraphics[width=0.12\textwidth]{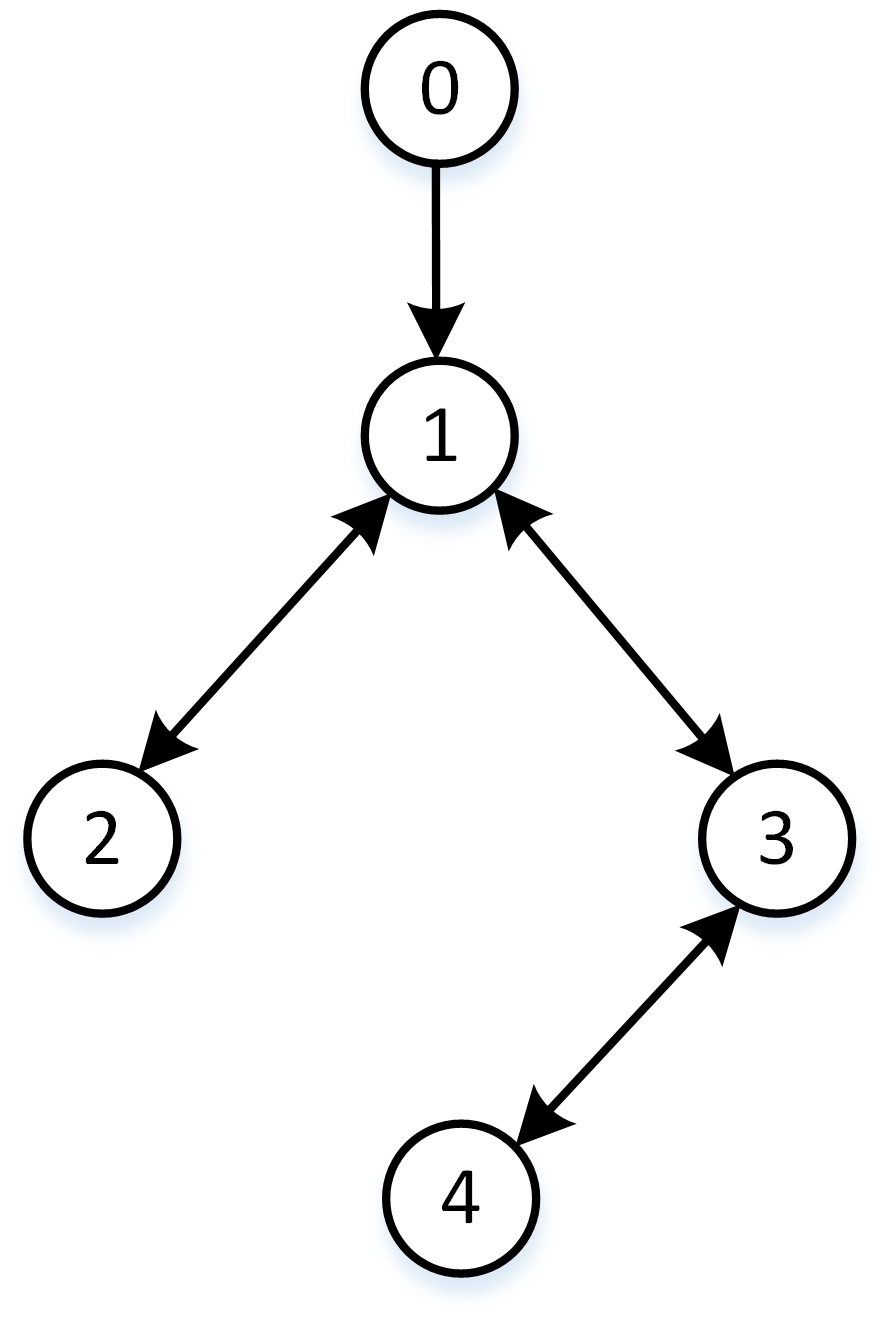}%
        \label{fig:a}%
        }%
    \subfloat[]{%
        \includegraphics[width=0.12\textwidth]{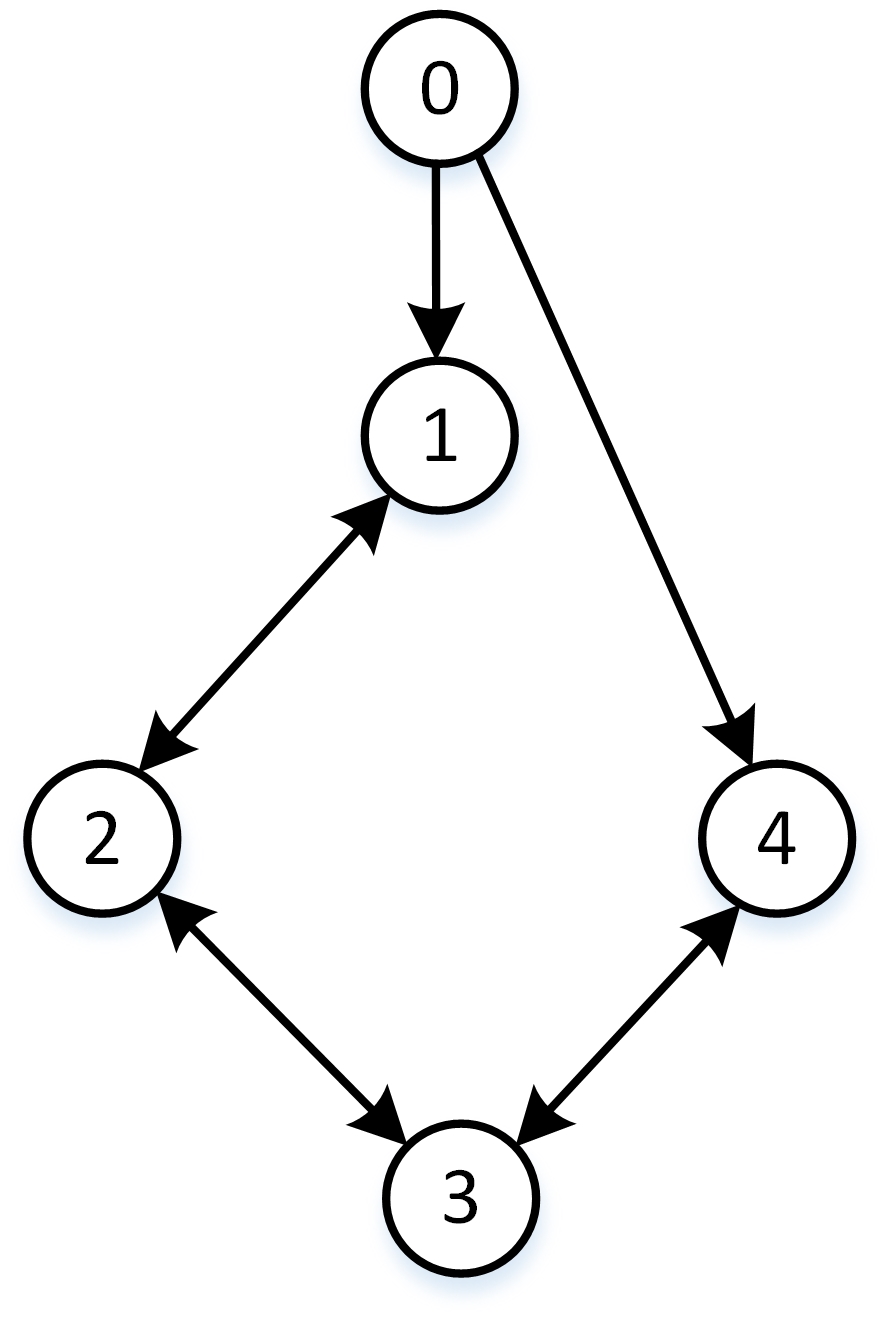}%
        \label{fig:b}%
        }%
    \subfloat[]{%
        \includegraphics[width=0.12\textwidth]{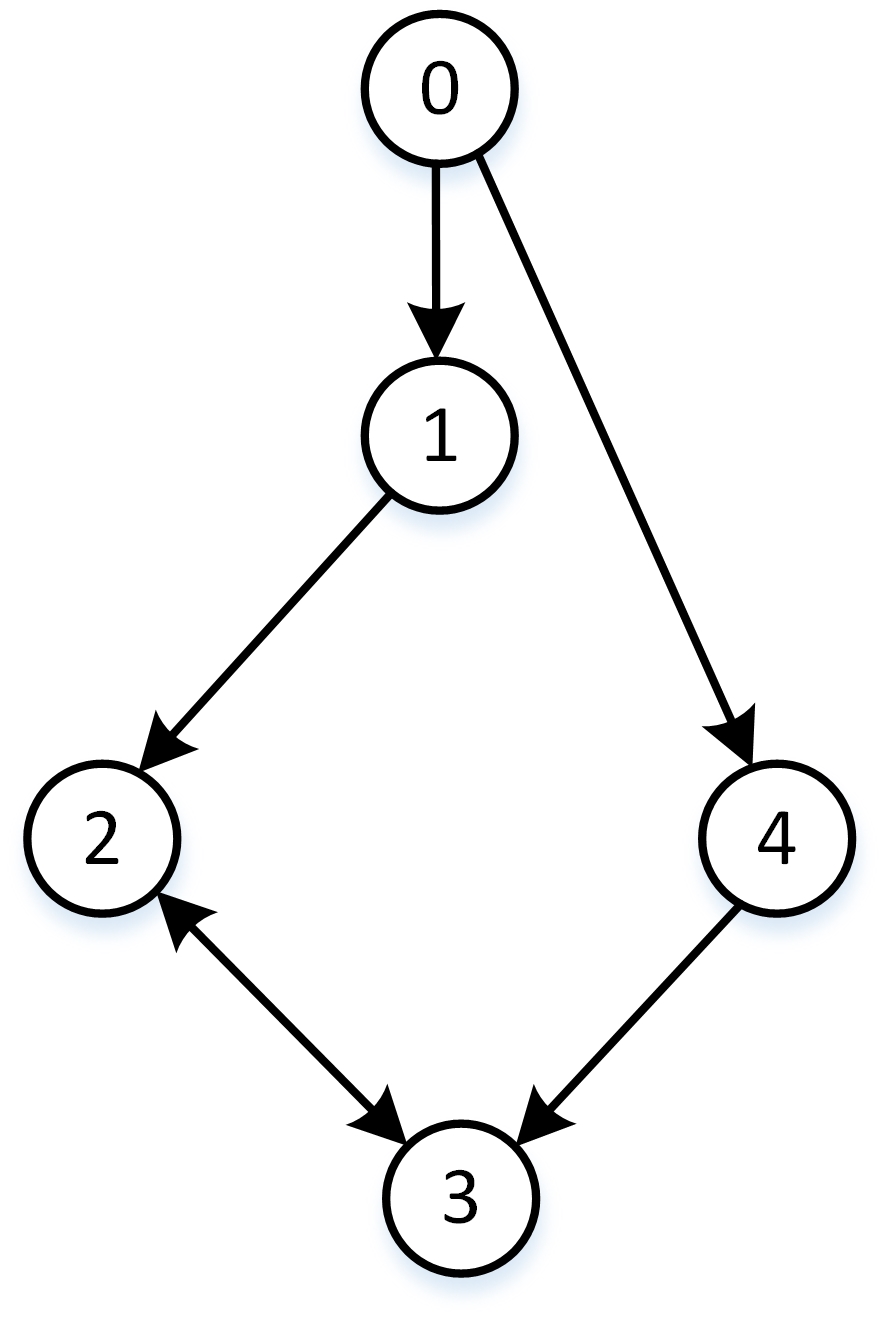}%
        \label{fig:c}%
        }%
    \caption{Topology of the communication graph in (a) Example 1, (b) Example 2, (c) Example 3.}\label{fig:Topology}
\end{figure}

\begin{figure}
    \centering
    \subfloat[]{%
        \includegraphics[width=0.22\textwidth]{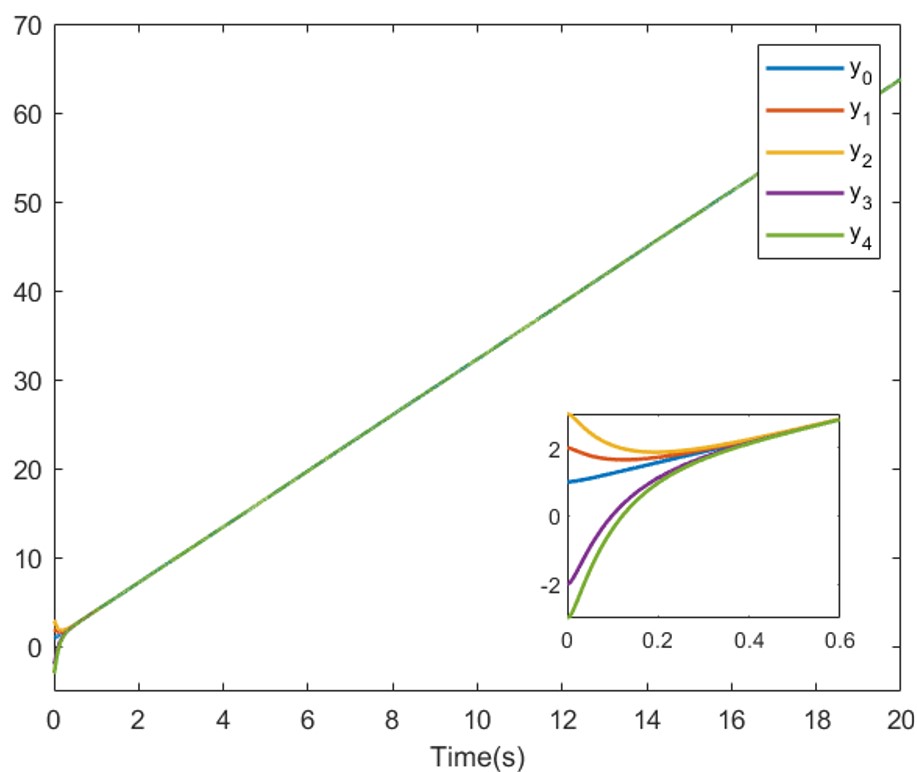}%
        \label{fig:a2}%
        }%
    \subfloat[]{%
        \includegraphics[width=0.22\textwidth]{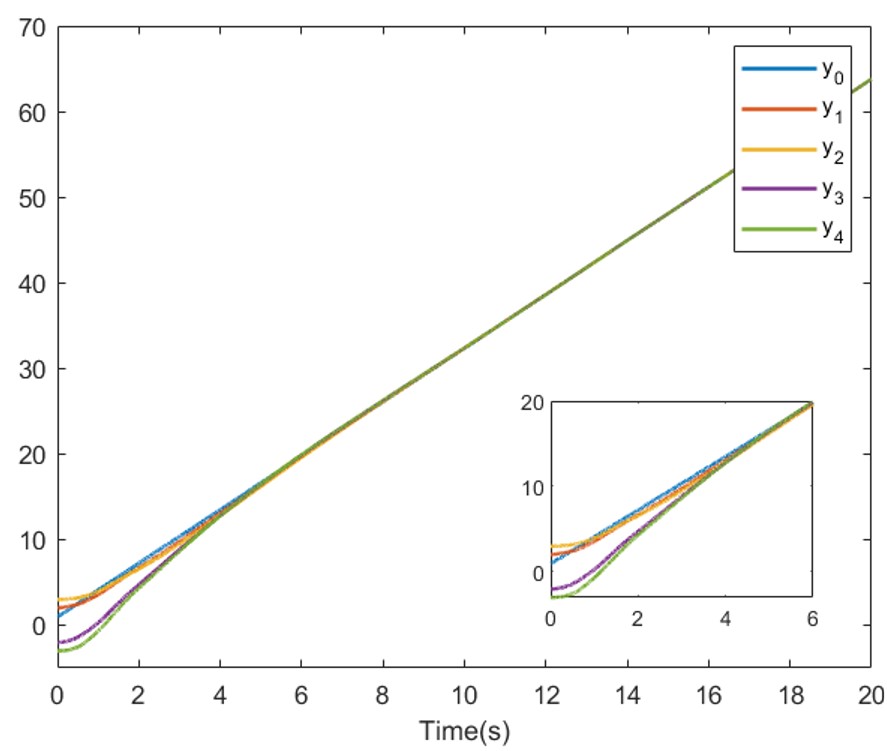}%
        \label{fig:b2}%
        }%
    \caption{Comparison of consensus tracking performance by the proposed method with unknown nonlinear dynamics (a) and the method in \cite{9906584} with known nonlinear dynamics (b) in Example $1$ with a passive leader.}
    \label{fig:ex1track1}
\end{figure}

The activation functions of RL Neural networks are selected as Gaussian functions $\exp \left(\frac{\left\|x-v_i\right\|^2}{\sigma_i^2}\right)$ with three neurons for each of the actor, critic, and estimator neural networks. The centers of the activation functions $(v_i)$ are uniformly chosen within $[-2,2]$. The width of the activation functions $\sigma_i$ is assumed to be 1. The gains 
$\Gamma_{c,ij}$, $\Gamma_{a,ij}$, $\Gamma_{\theta,ij}$ are all assumed equal to $10$, and the remaining gains of the adaptation laws are set equal to $1$. Moreover, $\gamma_{d,ij}=1$, $\sigma_{1d,ij}=2$, $k_{ij}=50$, $k_{p,ij}=1$, and $k_{q,ij}=1$.
First, consider a passive leader with $(u_0(t)=0)$. The results of the proposed method and comparison with \cite{9906584} are illustrated in Figures \ref{fig:ex1track1} and \ref{fig:ex1weightacstep1}. Figure \ref{fig:ex1track1} presents the output signals of both the leader and the followers. In the proposed method, the followers align with the leader in less than $0.4$ seconds, a significant improvement over the approximate $6$ seconds required in the method presented in \cite{9906584}. 
Figure \ref{fig:ex1weightacstep1} displays the norm of the weight vectors of the neural network of the critic and the actor for the first step in our proposed method.


\begin{figure}
    \centering
    \includegraphics[width=1\linewidth]{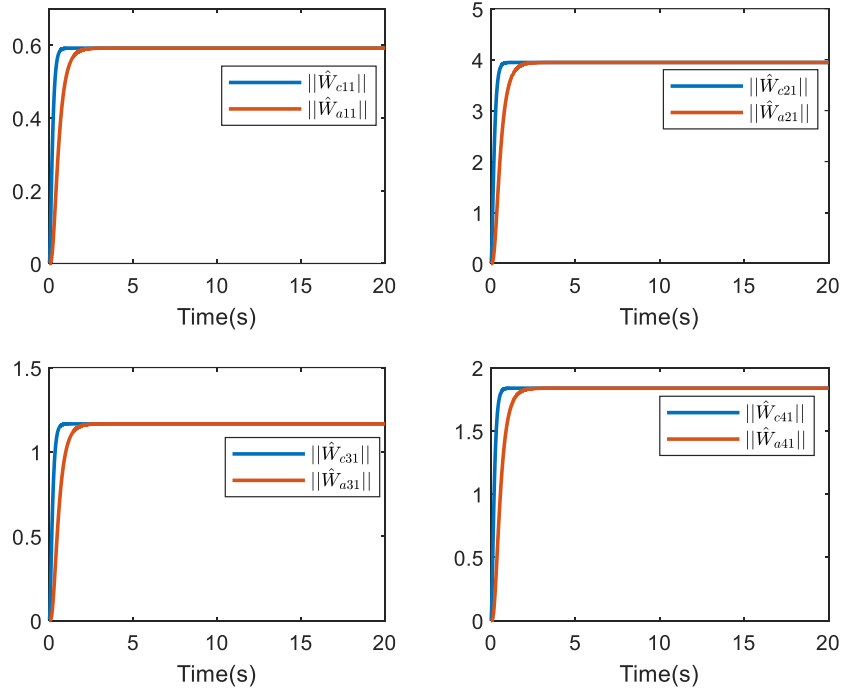}
    \caption{Weights of critic and actor neural networks of step $1$ of the proposed method in
    Example $1$ with a passive leader.}
    \label{fig:ex1weightacstep1}
\end{figure}

In the second scenario, the leader is active with a control input that remains inaccessible to the
followers. Figure \ref{fig:ex1trackactive1} shows the leader's and followers' positions for our proposed
method and \cite{9906584}. The leader's reference path is assumed to follow $10 \sin \left(\frac{2 \pi}{5} t\right)$. Figure \ref{fig:ex1trackactive2} illustrates the tracking error associated with both methods. Figures \ref{fig:ex1trackactive1} and \ref{fig:ex1trackactive2} collectively demonstrate the superior performance of our proposed method, showing faster convergence and reduced tracking error compared to \cite{9906584}. Figure \ref{fig:ex1weight1active}  displays the norm of the vector weights of the critic and actor neural networks for the first step of our proposed method, respectively.
\begin{figure}
    \centering
    \subfloat[]{%
        \includegraphics[width=0.22\textwidth]{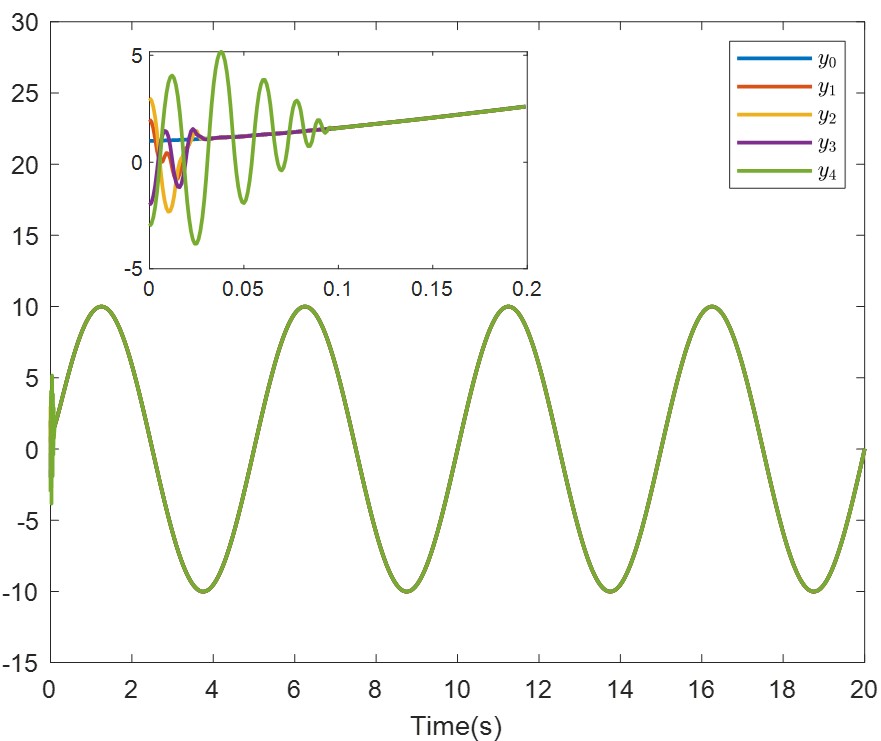}%
        \label{fig:a3}%
        }%
    \subfloat[]{%
        \includegraphics[width=0.22\textwidth]{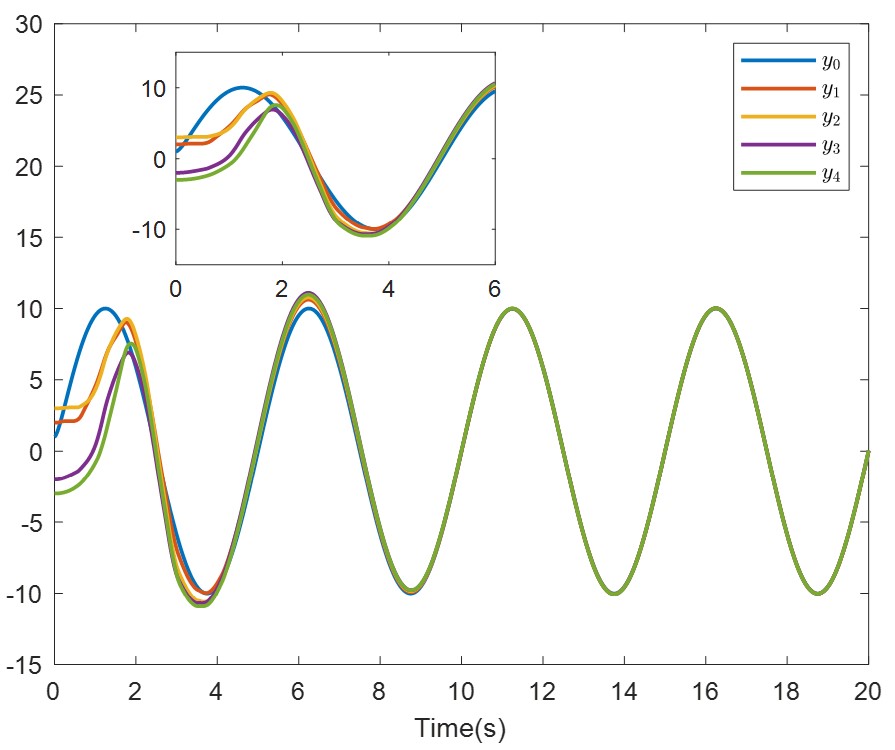}%
        \label{fig:b3}%
        }%
    \caption{Consensus tracking performance by the proposed method with unknown nonlinear dynamics (a) and \cite{9906584} with known nonlinear dynamics (b) in Example $1$ with an active leader.}
    \label{fig:ex1trackactive1}
\end{figure}

\begin{figure}
    \centering
    \subfloat[]{%
        \includegraphics[width=0.22\textwidth]{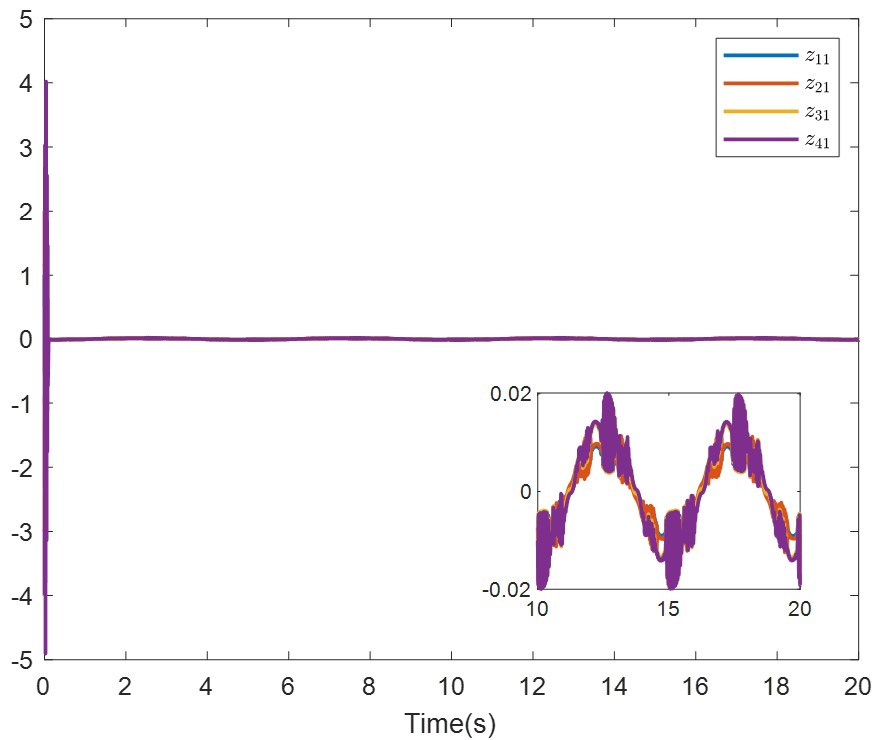}%
        \label{fig:a4}%
        }%
    \subfloat[]{%
        \includegraphics[width=0.22\textwidth]{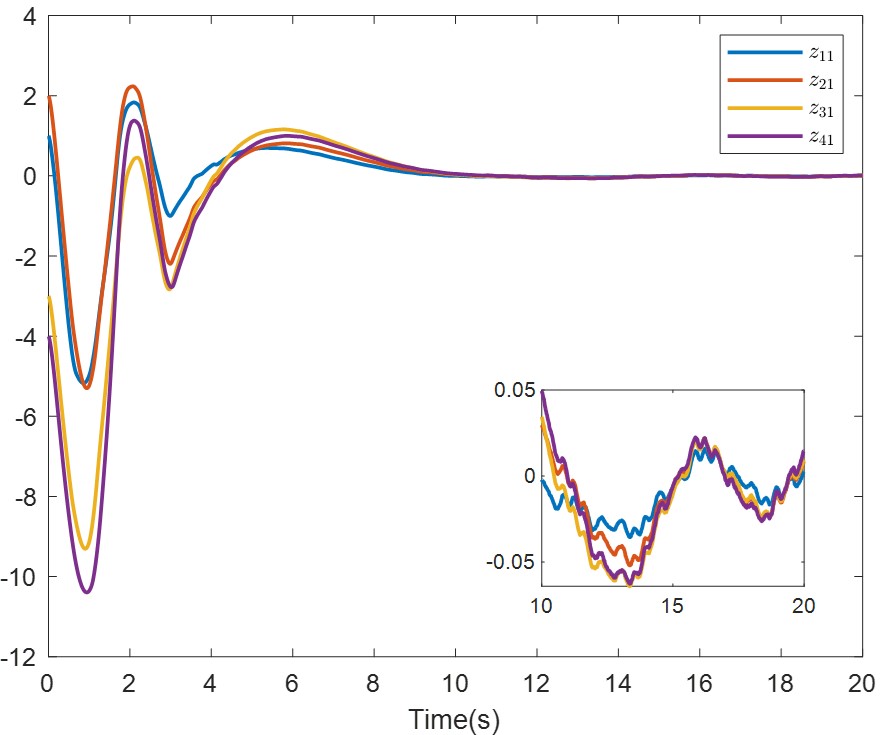}%
        \label{fig:b4}%
        }%
    \caption{Consensus tracking error by the proposed method (a) and \cite{9906584} (b) in Example $1$ with an active leader.}
    \label{fig:ex1trackactive2}
\end{figure}

\begin{figure}
    \centering
    \includegraphics[width=.9\linewidth]{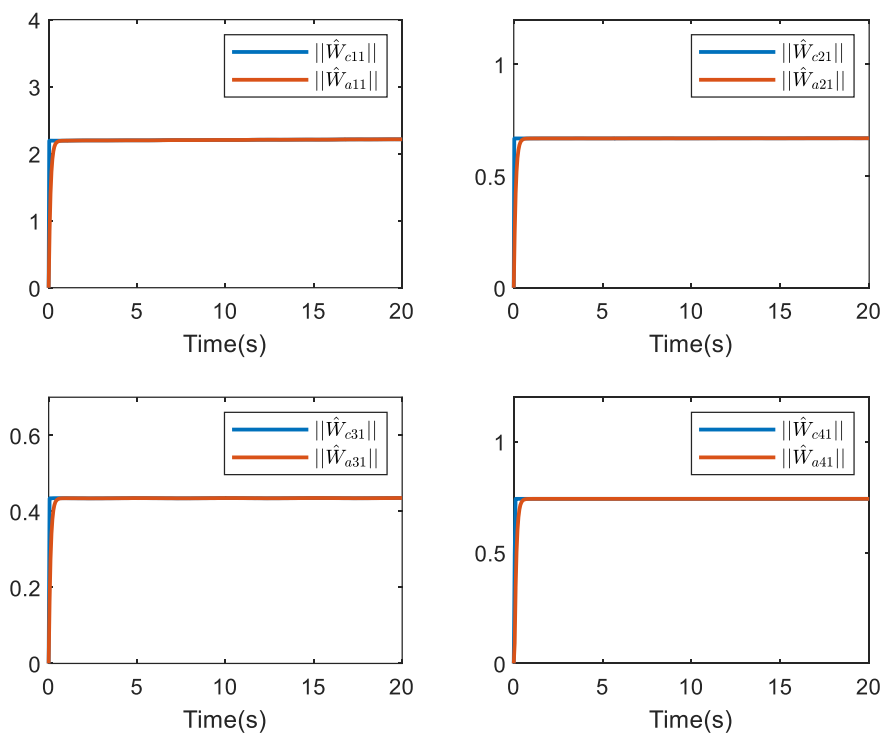}
    \caption{Weights of critic and actor 
     neural networks of step $1$ of the proposed method in Example $1$ with an active leader.}
    \label{fig:ex1weight1active}
\end{figure}

\begin{figure}
    \centering
    \subfloat[]{%
        \includegraphics[width=0.2\textwidth]{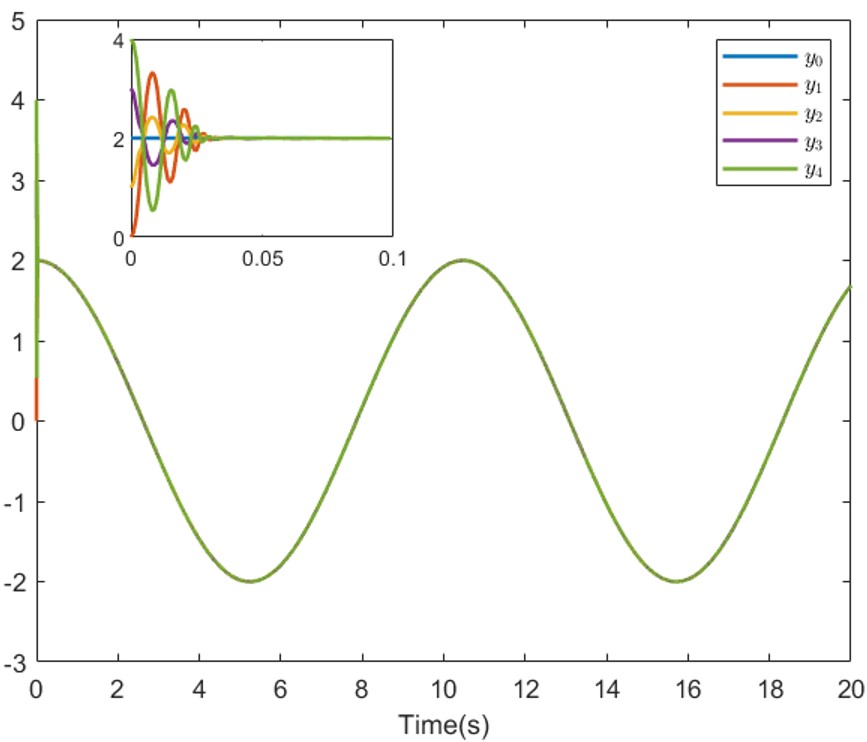}%
        \label{fig:a1}%
        }%
    \subfloat[]{%
        \includegraphics[width=0.2\textwidth]{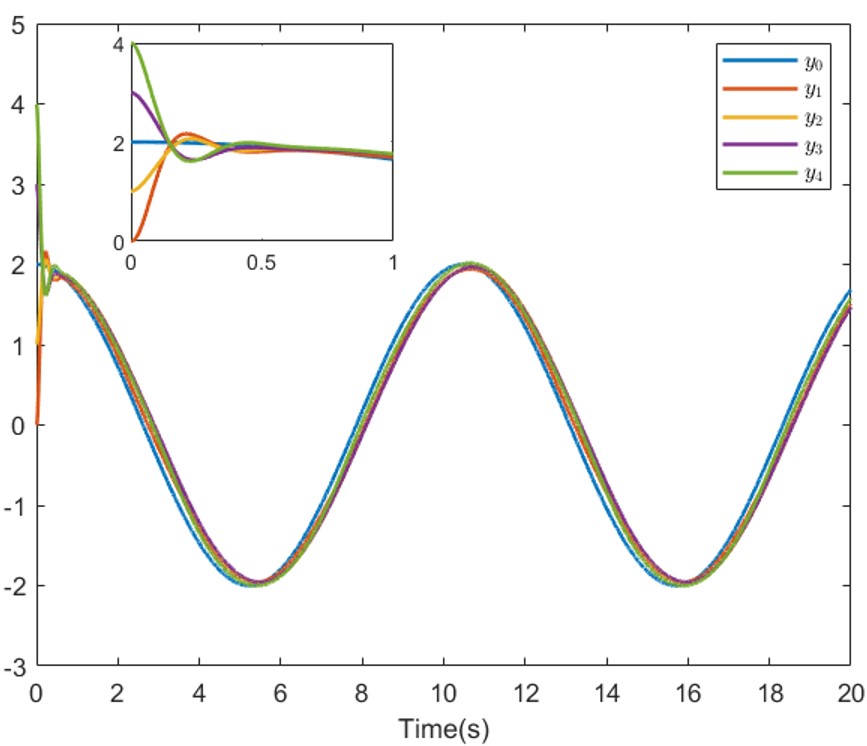}%
        \label{fig:b1}%
        }%
    \caption{Comparison of consensus tracking performance by the proposed method (a) and  \cite{9525047} (b) in Example $2$}
    \label{fig:ex2track1}
\end{figure}

\begin{figure}
    \centering
    \includegraphics[width=.9\linewidth]{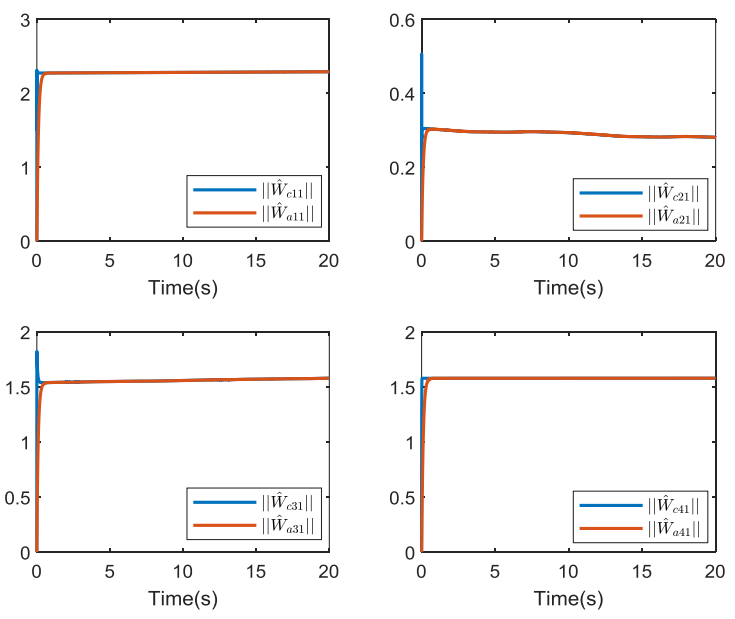}
    \caption{Weights of the critic and actor neural networks for step $1$ in Example $2$.}
    \label{fig:ex2weightacstep1}
\end{figure}

\begin{figure}
    \centering
    \includegraphics[width=.65\linewidth]{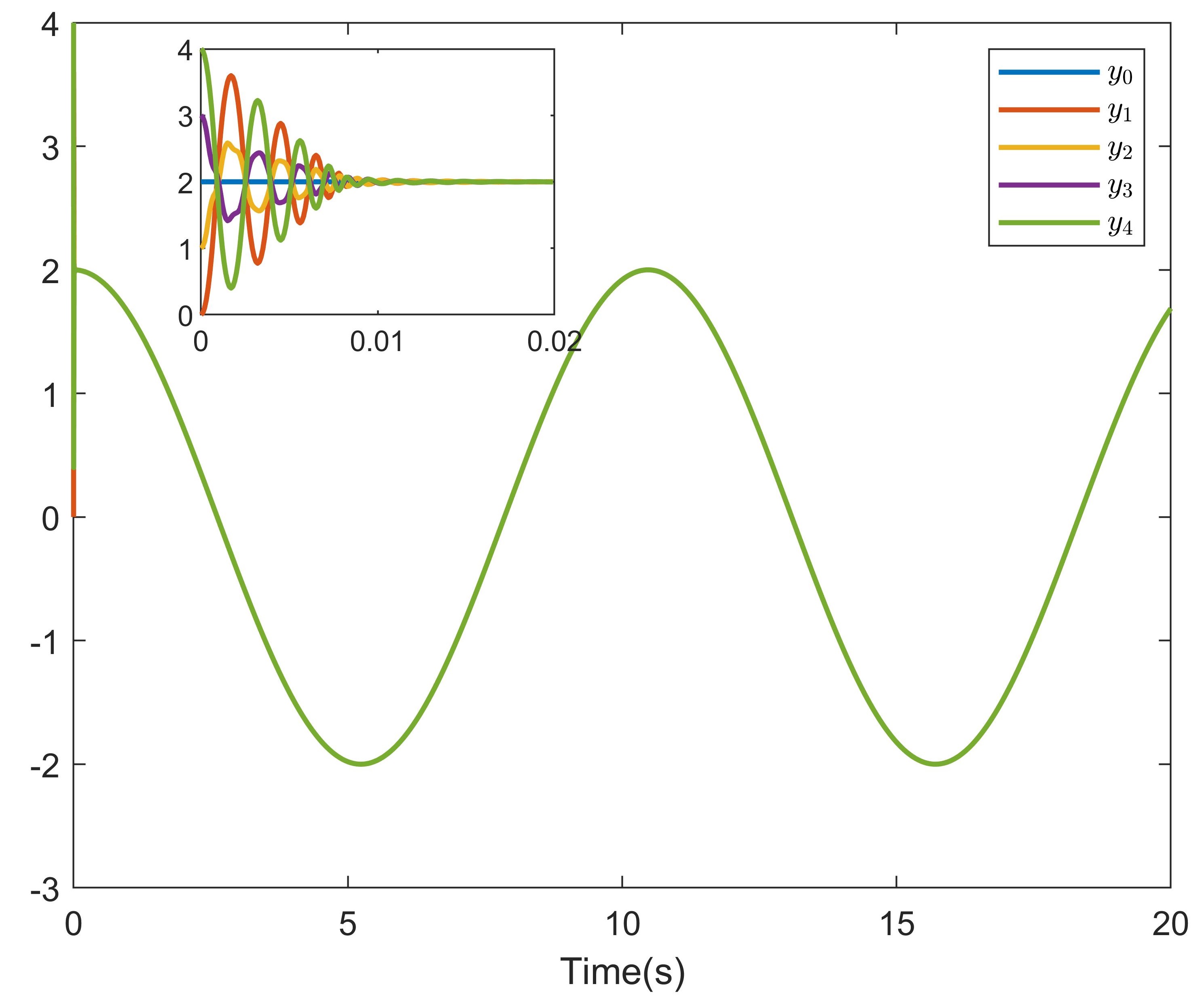}
    \caption{Consensus tracking performance by the proposed method in Example $3$}
    \label{fig:ex3track1}
\end{figure}

\begin{figure}
    \centering
    \includegraphics[width=0.9\linewidth]{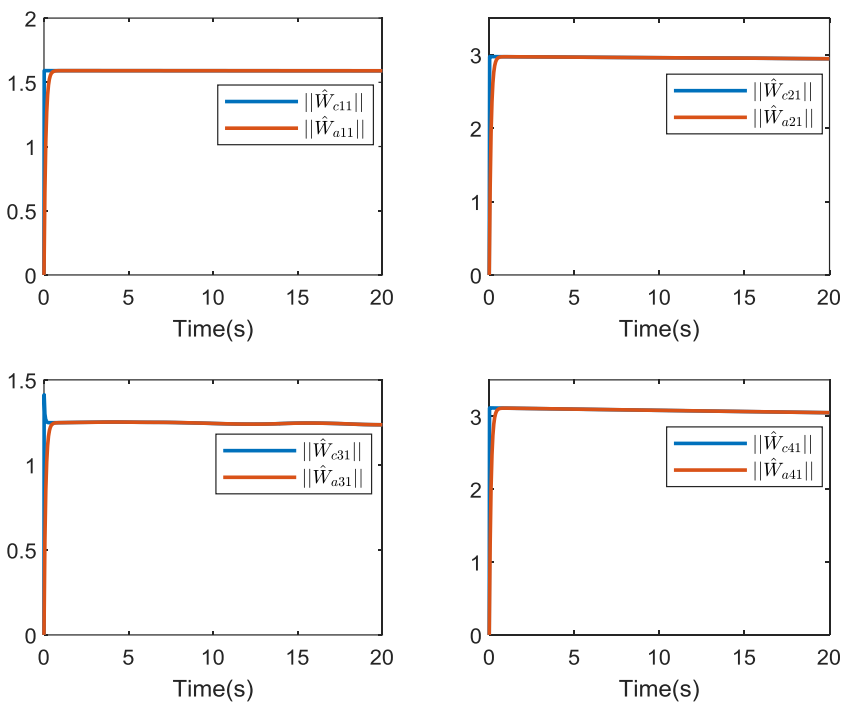}
    \caption{Norm of the weights of the critic and actor neural networks for step $1$ in Example $3$}
    \label{fig:ex3weightacstep1}
\end{figure}

 \end{example}

\begin{example}\label{ex:2}
Consider a leader-follower scheme with communication graph topology given in Figure \ref{fig:Topology}(b). The dynamics of  agents ($i=1,2,3,4$) are given by,
\begin{equation}
    \begin{aligned}
    &\dot{x}_{i 1}(t)=x_{i 2}-a_i \cos ^2\left(x_{i 1}\right)+b_i \sin \left(x_{i 1}\right) \\
    &\dot{x}_{i 2}(t)=u_i-c_i x_{i 2} \sin \left(x_{i 1}\right)+d_i \cos \left(x_{i 2}\right),
\end{aligned}
\end{equation}
where $a_i = [1.5,\! -0.8,\! 0.6,\! -1.3]$, $b_i = [-0.8,\! 0.4,\! -0.7,\! 0.8]$, $c_i = [0.7,\! 1.4,\! -1.5,\! -1.2]$, and $d_i = [0.5,\! -0.6,\! 1.1,\! -1.9]$~\cite{9525047}.
According to the restrictions of the problem formulation in
\cite{9525047}, disturbances are ignored, and the communication graph is assumed to be undirected. The leader output is $y_r (t)=2cos(0.6t)$, which is a reference signal for the followers. 
Three neurons are considered for the actor and critic neural network with Gaussian activation functions.The centers $(v_i)$ are uniformly chosen within $[-5,5]$. The width of the activation functions $\sigma_i$ is assumed to be $5$. The gains 
$\Gamma_{c,ij}$, $\Gamma_{a,ij}$, $\Gamma_{\theta,ij}$ are assumed equal to $15$, and the remaining gains of the adaptation laws are set equal to $1$. Moreover, $\gamma_{d,ij}=0.01$, $\sigma_{1d,ij}=1$, $k_{ij}=30$, $k_{p,ij}=1.5$, and $k_{q,ij}=1.5$.
The results of this example are illustrated in Figures \ref{fig:ex2track1} and \ref{fig:ex2weightacstep1}. Figure \ref{fig:ex2track1} illustrates the results of the followers, from which it can be inferred that the controller of our proposed method exhibits superior precision and transient response in following the leader compared to the method in\cite{9525047}. As demonstrated in Figure \ref{fig:ex2track1}, the tracking error of our proposed method is approximately one-tenth of the tracking error observed in the method \cite{9525047}. Figure \ref{fig:ex2weightacstep1} displays the norm of the weight vectors of the neural network of the critic and the actor for the first step in our proposed method.

\end{example}

\begin{example}\label{ex:3}
Consider a directed communication graph as depicted in Figure \ref{fig:Topology}(c).  The dynamics of the agents are the same as in Example $2$, while each subsystem now incorporates a disturbance, where the disturbance of the first layer is taken as $\sin(t)$ and the disturbance of the second layer is taken as $\cos(0.5t)$. The leader output, controller coefficients, neural
network parameters, and adaptation laws are chosen in alignment with Example $2$. The simulation
results for the the proposed method are shown in Figures \ref{fig:ex3track1} and \ref{fig:ex3weightacstep1}.
Figure \ref{fig:ex3track1} presents the leader's output along with the followers' output. It demonstrates the robust tracking performance of the proposed method, despite the
presence of uncertainties and external disturbances. Figure  \ref{fig:ex3weightacstep1}  depicts the norm of the weights of the critic and actor neural networks for the first step of the proposed algorithm.
\end{example}

\section{Conclusion}\label{sec:6}
This paper proposes a distributed RL-based backstepping controller to achieve fixed-time consensus in multi-agent systems with strict-feedback dynamics. The adaptation laws for the RL actor-critic neural networks, along with adaptive estimator networks for model uncertainties and external disturbances, are systematically developed to guarantee fixed-time consensus. The control algorithm decomposes the follower dynamics into a sequence of interconnected subsystems, with control policies designed iteratively for each layer. Beginning with the innermost subsystem, an RL-based control policy is constructed. This policy is then recursively incorporated into the next outer subsystem, continuing until the entire system is stabilized. The central idea is to ensure that the fixed-time stability achieved at each step contributes constructively to the global fixed-time consensus of the entire multi-agent system. Simulation results validate the performance of the proposed approach, demonstrating its capability to achieve rapid consensus while optimizing control effort. Future work may explore the extension of RL-based backstepping strategies to scenarios involving communication delays and time-varying interaction topologies.

\section*{Appendix}


\subsection*{Appendix A: Proof of Lemma 4}
\textbf{Proof:} Let the vectors $a$ and $b$ be given as $a=\left[a_1, a_2, \cdots, a_n\right]^\top, b=\left[b_1, b_2, \cdots, b_n\right]^\top$. Consider the vector multiplication,
\begin{equation*}
    -a^\top(a+b)^{\frac{1}{3}}=-\sum_{i=1}^n a_i\left(a_i+b_i\right)^{\frac{1}{3}}.
\end{equation*}
Using Young's inequality (Lemma 3),
  \begin{equation}
      -a_i^3 b_i \leq \frac{3}{4} a_i^4+\frac{1}{4} b_i^4 \leq \frac{7}{8} a_i^4+\frac{1}{4} b_i^4,
      \tag{A-1}
\end{equation}
  \begin{equation}
     \frac{3}{2} a_i^{\frac{4}{3}} b_i^{\frac{8}{3}} \leq \frac{3}{4} a_i^{\frac{8}{3}} b_i^{\frac{4}{3}}+\frac{3}{4} b_i^4.
     \tag{A-2}
 \end{equation}
Applying (A-1) and (A-2) results in,
\begin{align*}
   & -a_i^4-a_i^3 b_i \leq-\frac{1}{8} a_i^4+\frac{3}{4} a_i^{\frac{8}{3}}b_i^{\frac{4}{3}}-\frac{3}{2} a_i^{\frac{4}{3}} b_i^{\frac{8}{3}}+b_i^4,
\end{align*}
or equivalently,
\[
-a_i^3(a_i + b_i) \leq \left( -\frac{1}{2} a_i^{\frac{4}{3}} + b_i^{\frac{4}{3}} \right)^3.
\]
Taking cube roots, we get, $-a_i (a_i + b_i)^{\frac{1}{3}} \leq -\frac{1}{2} a_i^{\frac{4}{3}} + b_i^{\frac{4}{3}}.$ Summing over all \( i = 1, \dots, n \), it follows that,
\[
-\sum_{i=1}^n a_i (a_i + b_i)^{\frac{1}{3}} \leq \sum_{i=1}^n \left( -\frac{1}{2} a_i^{\frac{4}{3}} + b_i^{\frac{4}{3}} \right). \qed
\]

\subsection*{Appendix B: Proof of Lemma 5}
\textbf{Proof:} Consider the vector multiplication,
\begin{equation*}
    -a^\top(a+b)^3=-\sum_{i=1}^n a_i\left(a_i+b_i\right)^3= -a_i^4-a_i b_i^3-3 a_i^3 b_i-3 a_i^2 b_i^2.
\end{equation*}
Using Young's inequality,
\begin{equation}
    -a_i b_i^3 \leq \frac{1}{4} a_i^4+\frac{3}{4} b_i^4,
    \tag{B-1}
\end{equation}
\begin{equation}
    -3 a_i^3 b_i=\left(\frac{a_i}{\alpha}\right)^3\left(-3 \alpha^3 b_i\right) \leq \frac{3}{4 \alpha^4} a_i^4+\frac{81}{4} \alpha^{12} b_i^4.
    \tag{B-2}
\end{equation}

Let $\alpha^4=2$. Then $-3 a_i^3 b_i \leq \frac{3}{8} a_i^4+162 b_i^4$. Moreover,

\begin{equation}
    -3 a_i^2 b_i^2=\frac{a_i^2}{\sqrt{2}}\left(-3 \sqrt{2} b_i^2\right) \leq \frac{a_i^4}{4}+9 b_i^4.
    \tag{B-3}
\end{equation}

Applying (B-1), (B-2), and (B-3) yields,
\begin{align*}
  & -a^\top(a+b)^3 \leq -\frac{1}{8} \sum_{i=1}^n a_i^4 + 172 \sum_{i=1}^n b_i^4. \qed
\end{align*}

\subsection*{Appendix C: Proof of Lemma 9}
The time derivative of the Lyapunov function introduced in \eqref{eq:lyap1} is upper-bounded by,

\begin{equation}
\begin{aligned}
&\dot{V}_{i1} = 
 -k_{i1} e_i^2 - k_{p,i1} e_i^{p+1} - k_{q,i1} e_i^{q+1} \\
&- e_i \hat{W}_{a,i1}^\top S_{i1}(e_i) 
+ e_i \delta_{i1}(X_{i1}) \\
& - \sigma_{1c,i1} \tilde{W}_{c,i1}^\top S_{i1}(e_i) S_{i1}^\top(e_i) \hat{W}_{c,i1} 
- \sigma_{2c,i1} \tilde{W}_{c,i1}^\top \hat{W}_{c,i1}^p \\
&- \sigma_{3c,i1} \tilde{W}_{c,i1}^\top \hat{W}_{c,i1}^q 
- \tilde{W}_{c,i1}^\top S_{i1}(e_i) e_i \\
& - \sigma_{1a,i1} \tilde{W}_{a,i1}^\top S_{i1}(e_i) S_{i1}^\top(e_i)(\hat{W}_{a,i1} - \hat{W}_{c,i1}) \\
&- \sigma_{2a,i1} \tilde{W}_{a,i1}^\top (\hat{W}_{a,i1} - \hat{W}_{c,i1})^p \\
& - \sigma_{3a,i1} \tilde{W}_{a,i1}^\top (\hat{W}_{a,i1} - \hat{W}_{c,i1})^q 
- \sigma_{1\theta,i1} \tilde{\theta}_{i1}^\top \hat{\theta}_{i1}^p 
- \sigma_{2\theta,i1} \tilde{\theta}_{i1}^\top \hat{\theta}_{i1}^q \\
& - \sigma_{1D,i1} \tilde{D}_{i1} \hat{D}_{i1}^p 
- \sigma_{2D,i1} \tilde{D}_{i1} \hat{D}_{i1}^q 
- \frac{1}{\gamma_{D,i1}} \tilde{D}_{i1} \dot{D}_{i1}.
\end{aligned} \tag{C-1}
\end{equation}

First, note that $-e_i \hat{W}_{a,i1}^\top S_{i1}(e_i) 
- \tilde{W}_{c,i1}^\top S_{i1}(e_i) e_i = -e_i \tilde{W}_{a,i1}^\top S_{i1}(e_i) 
- e_i \hat{W}_{c,i1}^\top S_{i1}(e_i).$ Moreover, let $\bar{\delta}_{i 1}$ be the bound of the estimation error of the network for approximating unknown functions, then, 

\begin{align*}
& e_i \, \delta_{i1}(X_{i1}) 
\leq \tfrac{1}{2} e_i^2 + \tfrac{1}{2} \delta_{i1}^2(X_{i1}) 
\leq \tfrac{1}{2} e_i^2 + \tfrac{1}{2} \bar{\delta}_{i1}, \\
& -e_i \tilde{W}_{a,i1}^\top S_{i1}(e_i) 
\leq \tfrac{1}{2} e_i^2 
+ \tfrac{1}{2} \tilde{W}_{a,i1}^\top S_{i1}(e_i) S_{i1}^\top(e_i) \tilde{W}_{a,i1}, \\
& -e_i \hat{W}_{c,i1}^\top S_{i1}(e_i) 
\leq \tfrac{1}{2} e_i^2 
+ \tfrac{1}{2} \hat{W}_{c,i1}^\top S_{i1}(e_i) S_{i1}^\top(e_i) \hat{W}_{c,i1}, \\
& -\sigma_{1c,i1} \tilde{W}_{c,i1}^\top S_{i1}(e_i) S_{i1}^\top(e_i) \hat{W}_{c,i1} \\
& = -\tfrac{\sigma_{1c,i1}}{2} \tilde{W}_{c,i1}^\top S_{i1}(e_i) S_{i1}^\top(e_i) \tilde{W}_{c,i1} \\
& \quad -\tfrac{\sigma_{1c,i1}}{2} \hat{W}_{c,i1}^\top S_{i1}(e_i) S_{i1}^\top(e_i) \hat{W}_{c,i1} \\
& \quad +\tfrac{\sigma_{1c,i1}}{2} {W_{i1}^*}^\top S_{i1}(e_i) S_{i1}^\top(e_i) W_{i1}^*, \\
& \leq \tfrac{\sigma_{1c,i1}}{2} \tilde{W}_{c,i1}^\top S_{i1}(e_i) S_{i1}^\top(e_i) \tilde{W}_{c,i1} \\
& \quad -\tfrac{\sigma_{1c,i1}}{2} \hat{W}_{c,i1}^\top S_{i1}(e_i) S_{i1}^\top(e_i) \hat{W}_{c,i1} 
+ \tfrac{\sigma_{1c,i1}}{2} \bar{\lambda}_{s,i1} {W_{i1}^*}^\top W_{i1}^*,
\end{align*}
where $\bar{\lambda}_{s,i1}$ denotes the largest eigenvalue of $S_{i 1}\left(e_i\right) S_{i 1}^\top\left(e_i\right)$ in this context. Moreover,

\begin{align*}
&-\sigma_{2 c, i 1} \tilde{W}_{c, i 1}^\top \hat{W}_{c, i 1}^p 
= -\sigma_{2 c, i 1} \tilde{W}_{c, i 1}^\top \big(\tilde{W}_{c, i 1} + W_{i 1}^*\big)^p \\
&\leq -\frac{\sigma_{2 c, i 1}}{2} \left\| \tilde{W}_{c, i 1}^{\frac{p+1}{2}} \right\|^2 
+ \sigma_{2 c, i 1} \left\| W_{i 1}^{*\frac{p+1}{2}} \right\|^2, \\[1em]
&-\sigma_{3 c, i 1} \tilde{W}_{c, i 1}^\top \hat{W}_{c, i 1}^q 
= -\sigma_{3 c, i 1} \tilde{W}_{c, i 1}^\top \big(\tilde{W}_{c, i 1} + W_{i 1}^*\big)^q \\
&\leq -\underline{\sigma}_{3 c, i 1} \left\| \tilde{W}_{c, i 1}^{\frac{q+1}{2}} \right\|^2 
+ \bar{\sigma}_{3 c, i 1} \left\| W_{i 1}^{*\frac{q+1}{2}} \right\|^2,\\
& -\sigma_{1 a, i 1} \tilde{W}_{a, i 1}^\top S_{i 1}(e_i) S_{i 1}^\top(e_i) \left(\hat{W}_{a, i 1} - \hat{W}_{c, i 1}\right) \\
&= -\sigma_{1 a, i 1} \tilde{W}_{a, i 1}^\top S_{i 1}(e_i) S_{i 1}^\top(e_i) \tilde{W}_{a, i 1}\\
&+ \sigma_{1 a, i 1} \tilde{W}_{a, i 1}^\top S_{i 1}(e_i) S_{i 1}^\top(e_i) \tilde{W}_{c, i 1} \\
& \leq -\frac{\sigma_{1 a, i 1}}{2} \tilde{W}_{a, i 1}^\top S_{i 1}(e_i) S_{i 1}^\top(e_i) \tilde{W}_{a, i 1}\\
&+ \frac{\sigma_{1 a, i 1}}{2} \tilde{W}_{c, i 1}^\top S_{i 1}(e_i) S_{i 1}^\top(e_i) \tilde{W}_{c, i 1},
\end{align*}

\begin{align*}
& -\sigma_{2 a, i 1} \tilde{W}_{a, i 1}^\top \left( \hat{W}_{a, i 1} - \hat{W}_{c, i 1} \right)^p 
= -\sigma_{2 a, i 1} \tilde{W}_{a, i 1}^\top \left( \tilde{W}_{a, i 1} - \tilde{W}_{c, i 1} \right)^p \\
& \leq -\frac{\sigma_{2 a, i 1}}{2} \left\| \tilde{W}_{a, i 1}^{\frac{p+1}{2}} \right\|^2 
+ \sigma_{2 a, i 1} \left\| \tilde{W}_{c, i 1}^{\frac{p+1}{2}} \right\|^2, \\[1ex]
& -\sigma_{3 a, i 1} \tilde{W}_{a, i 1}^\top \left( \hat{W}_{a, i 1} - \hat{W}_{c, i 1} \right)^q \\
&\leq -\underline{\sigma}_{3 a, i 1} \left\| \tilde{W}_{a, i 1}^{\frac{q+1}{2}} \right\|^2 
+ \bar{\sigma}_{3 a, i 1} \left\| \tilde{W}_{c, i 1}^{\frac{q+1}{2}} \right\|^2, \\[1ex]
& -\sigma_{1 \theta, i 1} \tilde{\theta}_{i 1}^\top \hat{\theta}_{i 1}^p 
\leq -\frac{\sigma_{1 \theta, i 1}}{2} \left\| \tilde{\theta}_{i 1}^{\frac{p+1}{2}} \right\|^2 
+ \sigma_{1 \theta, i 1} \left\| {\theta}_{i 1}^*{}^{\frac{p+1}{2}} \right\|^2, \\[1ex]
& -\sigma_{2 \theta, i 1} \tilde{\theta}_{i 1}^\top \hat{\theta}_{i 1}^q 
\leq -\underline{\sigma}_{2 \theta, i 1} \left\| \tilde{\theta}_{i 1}^{\frac{q+1}{2}} \right\|^2 
+ \bar{\sigma}_{2 \theta, i 1} \left\| {\theta}_{i 1}^*{}^{\frac{q+1}{2}} \right\|^2, \\[1ex]
& -\sigma_{1 D, i 1} \tilde{D}_{i 1} \hat{D}_{i 1}^p 
\leq -\frac{\sigma_{1 D, i 1}}{2} \tilde{D}_{i 1}^{p+1} + \sigma_{1 D, i 1} \bar{D}_{i 1}^{p+1}, \\[1ex]
& -\sigma_{2 D, i 1} \tilde{D}_{i 1} \hat{D}_{i 1}^q 
\leq -\underline{\sigma}_{2 D, i 1} \tilde{D}_{i 1}^{q+1} + \bar{\sigma}_{2 D, i 1} \bar{D}_{i 1}^{q+1}, \\[1ex]
& -\frac{1}{\gamma_{D, i 1}} \tilde{D}_{i 1} \dot{D}_{i 1} 
= -\left( \frac{1}{\mu_{D, i 1}} \tilde{D}_{i 1} \right) \left( \frac{\mu_{D, i 1}}{\gamma_{D, i 1}} \dot{D}_{i 1} \right) \\
& \leq \frac{1}{4 \mu_{D, i 1}^4} \tilde{D}_{i 1}^4 + \frac{3}{4} \left( \frac{\mu_{D, i 1} \dot{D}_{i 1}}{\gamma_{D, i 1}} \right)^{\frac{4}{3}}\\ 
&\leq \frac{1}{4 \mu_{D, i 1}^4} \tilde{D}_{i 1}^{q+1} + \frac{3}{4} \left( \frac{\mu_{D, i 1}}{\gamma_{D, i 1}} \right)^{\frac{4}{3}} \bar{D}_{d, i 1}^{\frac{4}{3}}.
\end{align*}

Substituting the derived inequalities in (C-1) gives,

\begin{equation*}
\begin{aligned}
&\dot{V}_{i1} =\;  -\left(k_{i1} - \frac{3}{2} \right) e_i^2 
- k_{p,i1} e_i^{p+1} 
- k_{q,i1} e_i^{q+1} 
+ \frac{1}{2} \bar{\delta}_{i1} \\
& -\left( \frac{\sigma_{1a,i1}}{2} - \frac{1}{2} \right) 
\tilde{W}_{a,i1}^\top S_{i1}(e_i) S_{i1}^\top(e_i) \tilde{W}_{a,i1} \\
& -\left( \frac{\sigma_{1c,i1}}{2} - \frac{1}{2} \right) 
\hat{W}_{c,i1}^\top S_{i1}(e_i) S_{i1}^\top(e_i) \hat{W}_{c,i1} \\
& -\left( \frac{\sigma_{1c,i1}}{2} - \frac{\sigma_{1a,i1}}{2} \right) 
\tilde{W}_{c,i1}^\top S_{i1}(e_i) S_{i1}^\top(e_i) \tilde{W}_{c,i1} \\
& + \frac{\sigma_{1c,i1}}{2} \bar{\lambda}_{s,i1} \left\| W_{i1}^* \right\|^2 -\left( \frac{\sigma_{2c,i1}}{2} - \sigma_{2a,i1} \right) 
\left\| \tilde{W}_{c,i1}^{\frac{p+1}{2}} \right\|^2\\
&  
+ \sigma_{2c,i1} \left\| {W_{i1}^*}^{\frac{p+1}{2}} \right\|^2  -\left( \underline{\sigma}_{3c,i1} - \bar{\sigma}_{3a,i1} \right) 
\left\| \tilde{W}_{c,i1}^{\frac{q+1}{2}} \right\|^2\\
& + \bar{\sigma}_{3c,i1} \left\| {W_{i1}^*}^{\frac{q+1}{2}} \right\|^2 
- \frac{\sigma_{2a,i1}}{2} \left\| \tilde{W}_{a,i1}^{\frac{p+1}{2}} \right\|^2 
- \underline{\sigma}_{3a,i1} \left\| \tilde{W}_{a,i1}^{\frac{q+1}{2}} \right\|^2 \\
& - \frac{\sigma_{1\theta,i1}}{2} \left\| \tilde{\theta}_{i1}^{\frac{p+1}{2}} \right\|^2 
+ \sigma_{1\theta,i1} \left\| {\theta_{i1}^*}^{\frac{p+1}{2}} \right\|^2 \\
&- \underline{\sigma}_{2\theta,i1} \left\| \tilde{\theta}_{i1}^{\frac{q+1}{2}} \right\|^2 
+ \bar{\sigma}_{2\theta,i1} \left\| {\theta_{i1}^*}^{\frac{q+1}{2}} \right\|^2 \\
& - \frac{\sigma_{1D,i1}}{2} \tilde{D}_{i1}^{p+1} 
+ \sigma_{1D,i1} \bar{D}_{i1}^{p+1} 
- \underline{\sigma}_{2D,i1} \tilde{D}_{i1}^{q+1} 
+ \bar{\sigma}_{2D,i1} \bar{D}_{i1}^{q+1} \\
& + \frac{1}{4 \mu_{D,i1}^4} \tilde{D}_{i1}^{q+1} 
+ \frac{3}{4} \left( \frac{\mu_{D,i1}}{\gamma_{D,i1}} \right)^{\frac{4}{3}} \bar{D}_{d,i1}^{\frac{4}{3}} 
+ g_i z_{i2} e_i.
\end{aligned}
\tag{C-2}
\end{equation*}

The bounds on the design parameters specified in Lemma 9 (see \eqref{eq:lemma9_parameters}) ensure that the Lyapunov function satisfies the following bound,

\begin{align*}
 &  \dot{V}_{i 1} \leq 
    -k_{p, i 1} \, e_i^{p+1} - k_{q, i 1} \, e_i^{q+1} -\sigma_{2 c, i 1} \left\| \tilde{W}_{c, i 1}^{\frac{p+1}{2}} \right\|^2 \\
   & 
     -\sigma_{3 c a, i 1} \left\| \tilde{W}_{c, i 1}^{\frac{q+1}{2}} \right\|^2 -\frac{\sigma_{2 a, i 1}}{2} \left\| \tilde{W}_{a, i 1}^{\frac{p+1}{2}} \right\|^2 
     -\underline{\sigma}_{3 a, i 1} \left\| \tilde{W}_{a, i 1}^{\frac{q+1}{2}} \right\|^2  \\
   & -\frac{\sigma_{1 \theta, i 1}}{2} \left\| \tilde{\theta}_{i 1}^{\frac{p+1}{2}} \right\|^2 
     -\underline{\sigma}_{2 \theta, i 1} \left\| \tilde{\theta}_{i 1}^{\frac{q+1}{2}} \right\|^2 \\
   & -\frac{\sigma_{1 D, i 1}}{2} \, \tilde{D}_{i 1}^{p+1} 
     -\tilde{\sigma}_{2 D, i 1} \, \tilde{D}_{i 1}^{q+1} 
     + C_{i 1} + g_i z_{i 2} e_i.
\end{align*}
Which completes the proof of Lemma 9. \qed

\subsection*{Appendix D: Proof of Lemma 10}

The time derivative of the Lyapunov function introduced in \eqref{eq:Lyapj} is upper bounded by, 

\begin{equation*}
\begin{aligned}
& \dot{V}_{i j} =\; \dot{V}_{i(j-1)} 
- z_{i j} z_{i(j-1)}^r 
- k_{i j} z_{i j}^2 
- k_{p, i j} z_{i j}^{p+1} 
- k_{q, i j} z_{i j}^{q+1} \\
& - \hat{W}_{a, i j}^\top S_{i j}(z_{i j}) z_{i j} 
- \tilde{\theta}_{i j}^\top \Phi_{i j}(\chi_{i j}) z_{i j} 
- \tilde{d}_{i j} z_{i j} \\
& + z_{i j} \delta_{i j}(X_{i j}) 
+ z_{i j} z_{i(j+1)} 
- \tilde{W}_{c, i j}^\top S_{i j}(z_{i j}) z_{i j} \\
& - \sigma_{1 c, i j} \tilde{W}_{c, i j}^\top S_{i j}(z_{i j}) S_{i j}^\top(z_{i j}) \tilde{W}_{c, i j} 
- \sigma_{2 c, i j} \tilde{W}_{c, i j}^\top \hat{W}_{c, i j}^p 
 \\
& - \sigma_{3 c, i j} \tilde{W}_{c, i j}^\top \hat{W}_{c, i j}^q - \sigma_{1 a, i j} \tilde{W}_{a, i j}^\top S_{i j}(z_{i j}) S_{i j}^\top(z_{i j}) (\hat{W}_{a, i j} - \hat{W}_{c, i j}) \\
& - \sigma_{2 a, i j} \tilde{W}_{a, i j}^\top (\hat{W}_{a, i j} - \hat{W}_{c, i j})^p 
- \sigma_{3 a, i j} \tilde{W}_{a, i j}^\top (\hat{W}_{a, i j} - \hat{W}_{c, i j})^q \\
& + \tilde{\theta}_{i j}^\top \Phi_{i j}(\chi_{i j}) z_{i j} 
- \sigma_{1 \theta, i j} \tilde{\theta}_{i j}^\top \hat{\theta}_{i j}^p 
- \sigma_{2 \theta, i j} \tilde{\theta}_{i j}^\top \hat{\theta}_{i j}^q \\
& + \tilde{d}_{i j} z_{i j} 
- \sigma_{1 d, i j} \tilde{d}_{i j} \hat{d}_{i j}^p 
- \sigma_{2 d, i j} \tilde{d}_{i j} \hat{d}_{i j}^q 
- \frac{1}{\gamma_{d, i j}} \tilde{d}_{i j} \dot{d}_{i j}.
\end{aligned} \tag{D-1}
\end{equation*}

According to the definition of $\tilde{W}_{c,ij}$ and $\tilde{W}_{a,ij}$, $-z_{i j} \hat{W}_{a, i j}^\top S_{i j}(z_{i j}) - \tilde{W}_{c, i j}^\top S_{i j}(z_{i j}) z_{i j}  = -z_{i j} \tilde{W}_{a, i j}^\top S_{i j}(z_{i j}) - z_{i j} \hat{W}_{c, i j}^\top S_{i j}(z_{i j})$. Accordingly,

\begin{align*}
    & \sigma_{1 c, i j} \tilde{W}_{c, i j}^\top S_{i j}(z_{i j}) S_{i j}^\top(z_{i j}) \hat{W}_{c, i j} \\
    &= -\frac{\sigma_{1 c, i j}}{2} \tilde{W}_{c, i j}^\top S_{i j}(z_{i j}) S_{i j}^\top(z_{i j}) \tilde{W}_{c, i j} \\
    &\quad -\frac{\sigma_{1 c, i j}}{2} \hat{W}_{c, i j}^\top S_{i j}(z_{i j}) S_{i j}^\top(z_{i j}) \hat{W}_{c, i j} \\
    &\quad +\frac{\sigma_{1 c, i j}}{2} W_{i j}^{*\top} S_{i j}(z_{i j}) S_{i j}^\top(z_{i j}) W_{i j}^*,\\
     & -z_{i j} \tilde{W}_{a, i j}^\top S_{i j}(z_{i j}) 
 \leq \frac{1}{2} z_{i j}^2 + \frac{1}{2} \tilde{W}_{a, i j}^\top S_{i j}(z_{i j}) S_{i j}^\top(z_{i j}) \tilde{W}_{a, i j}, \\
 & -z_{i j} \hat{W}_{c, i j}^\top S_{i j}(z_{i j}) 
 \leq \frac{1}{2} z_{i j}^2 + \frac{1}{2} \hat{W}_{c, i j}^\top S_{i j}(z_{i j}) S_{i j}^\top(z_{i j}) \hat{W}_{c, i j},\\
  & z_{i j} \delta_{i j}(X_{i j}) 
 \leq \frac{1}{2} z_{i j}^2 + \frac{1}{2} \delta_{i j}^2(X_{i j}) 
 \leq \frac{1}{2} z_{i j}^2 + \frac{1}{2} \bar{\delta}_{i j}, \\
 & -\sigma_{2 c, i j} \tilde{W}_{c, i j}^\top \hat{W}_{c, i j}^p 
 \leq -\frac{\sigma_{2 c, i j}}{2} \left\| \tilde{W}_{c, i j}^{\left(\frac{p+1}{2}\right)} \right\|^2 
 + \sigma_{2 c, i j} \left\| W_{i j}^*{}^{\left(\frac{p+1}{2}\right)} \right\|^2, \\
 & -\sigma_{3 c, i j} \tilde{W}_{c, i j}^\top \hat{W}_{c, i j}^q 
 \leq -\underline{\sigma}_{3 c, i j} \left\| \tilde{W}_{c, i j}^{\left(\frac{q+1}{2}\right)} \right\|^2 
 + \bar{\sigma}_{3 c, i j} \left\| W_{i j}^*{}^{\left(\frac{q+1}{2}\right)} \right\|^2,\\
& \sigma_{1 a, i j} \tilde{W}_{a, i j}^\top S_{i j}(z_{i j}) S_{i j}^\top(z_{i j}) \left( \hat{W}_{a, i j} - \hat{W}_{c, i j} \right)\\
&= -\sigma_{1 a, i j} \tilde{W}_{a, i j}^\top S_{i j}(z_{i j}) S_{i j}^\top(z_{i j}) \tilde{W}_{a, i j} \\
&+ \sigma_{1 a, i j} \tilde{W}_{a, i j}^\top S_{i j}(z_{i j}) S_{i j}^\top(z_{i j}) \tilde{W}_{c, i j}
\end{align*}

\begin{align*}
&\leq -\frac{\sigma_{1 a, i j}}{2} \tilde{W}_{a, i j}^\top S_{i j}(z_{i j}) S_{i j}^\top(z_{i j}) \tilde{W}_{a, i j}\\
&+ \frac{\sigma_{1 a, i j}}{2} \tilde{W}_{c, i j}^\top S_{i j}(z_{i j}) S_{i j}^\top(z_{i j}) \tilde{W}_{c, i j}, \\[1ex]
& -\sigma_{2 a, i j} \tilde{W}_{a, i j}^\top \left( \hat{W}_{a, i j} - \hat{W}_{c, i j} \right)^p 
= -\sigma_{2 a, i j} \tilde{W}_{a, i j}^\top \left( \tilde{W}_{a, i j} - \tilde{W}_{c, i j} \right)^p \\
&\leq -\frac{\sigma_{2 a, i j}}{2} \left\| \tilde{W}_{a, i j}^{\frac{p+1}{2}} \right\|^2 
+ \sigma_{2 a, i j} \left\| \tilde{W}_{c, i j}^{\frac{p+1}{2}} \right\|^2, \\[1ex]
& -\sigma_{3 a, i j} \tilde{W}_{a, i j}^\top \left( \hat{W}_{a, i j} - \hat{W}_{c, i j} \right)^q \\
&\leq -\underline{\sigma}_{3 a, i j} \left\| \tilde{W}_{a, i j}^{\frac{q+1}{2}} \right\|^2 
+ \bar{\sigma}_{3 a, i j} \left\| \tilde{W}_{c, i j}^{\frac{q+1}{2}} \right\|^2, \\[1ex]
& -\sigma_{2 \theta, i j} \tilde{\theta}_{i j}^\top \hat{\theta}_{i j}^q 
\leq -\underline{\sigma}_{2 \theta, i j} \left\| \tilde{\theta}_{i j}^{\frac{q+1}{2}} \right\|^2 
+ \bar{\sigma}_{2 \theta, i j} \left\| {\theta}_{i j}^*{}^{\frac{q+1}{2}} \right\|^2, \\[1ex]
& -\sigma_{1 d, i j} \tilde{d}_{i j} \hat{d}_{i j}^p 
\leq -\frac{\sigma_{1 d, i j}}{2} \tilde{d}_{i j}^{p+1} + \sigma_{1 d, i j} \bar{d}_{i j}^{p+1}, \\[1ex]
& -\sigma_{2 d, i j} \tilde{d}_{i j} \hat{d}_{i j}^q 
\leq -\underline{\sigma}_{2 d, i j} \tilde{d}_{i j}^{q+1} + \bar{\sigma}_{2 d, i j} \bar{d}_{i j}^{q+1}, \\[1ex]
& -\frac{1}{\gamma_{d, i j}} \tilde{d}_{i j} \dot{d}_{i j} 
= -\left( \frac{1}{\mu_{d, i j}} \tilde{d}_{i j} \right) \left( \frac{\mu_{d, i j}}{\gamma_{d, i j}} \dot{d}_{i j} \right) \\
& \leq \frac{1}{4 \mu_{d, i j}^4} \tilde{d}_{i j}^{q+1} + \frac{3}{4} \left( \frac{\mu_{d, i j}}{\gamma_{d, i j}} \right)^{\frac{4}{3}} \bar{d}_{d, i j}^{\frac{4}{3}}.
\end{align*}

Substituting the derived inequalities into (D-1) gives,

\begin{equation}
\begin{aligned}
   \dot{V}_{ij} \leq\ 
   & \dot{V}_{i(j-1)} 
   - z_{ij} z_{i(j-1)}^r 
   - \left(k_{ij} - \tfrac{3}{2}\right) z_{ij}^2 
   - k_{p,ij} z_{ij}^{p+1} 
   - k_{q,ij} z_{ij}^{q+1} \\
   & - \left(\tfrac{\sigma_{1a,ij}}{2} - \tfrac{1}{2}\right) \tilde{W}_{a,ij}^\top S_{ij}(z_{ij}) S_{ij}^\top(z_{ij}) \tilde{W}_{a,ij} \\
   & - \left(\tfrac{\sigma_{1c,ij}}{2} - \tfrac{1}{2}\right) \hat{W}_{c,ij}^\top S_{ij}(z_{ij}) S_{ij}^\top(z_{ij}) \hat{W}_{c,ij} \\
   & - \left(\tfrac{\sigma_{1c,ij}}{2} - \tfrac{\sigma_{1a,ij}}{2}\right) \tilde{W}_{c,ij}^\top S_{ij}(z_{ij}) S_{ij}^\top(z_{ij}) \tilde{W}_{c,ij} \\
   & - \left(\tfrac{\sigma_{2c,ij}}{2} - \sigma_{2a,ij}\right) \left\| \tilde{W}_{c,ij}^{\frac{p+1}{2}} \right\|^2 
   - \left(\underline{\sigma}_{3c,ij} - \bar{\sigma}_{3a,ij}\right) \left\| \tilde{W}_{c,ij}^{\frac{q+1}{2}} \right\|^2 \\
   & - \tfrac{\sigma_{2a,ij}}{2} \left\| \tilde{W}_{a,ij}^{\frac{p+1}{2}} \right\|^2 
   - \underline{\sigma}_{3a,ij} \left\| \tilde{W}_{a,ij}^{\frac{q+1}{2}} \right\|^2\\
   & -\frac{\sigma_{1\theta,ij}}{2} \left\| \tilde{\theta}_{ij}^{\frac{p+1}{2}} \right\|^2 
   - \underline{\sigma}_{2\theta,ij} \left\| \tilde{\theta}_{ij}^{\frac{q+1}{2}} \right\|^2 \\
   & - \frac{\sigma_{1d,ij}}{2} \tilde{d}_{ij}^{p+1} 
   - \left( \underline{\sigma}_{2d,ij} - \frac{1}{4\mu_{d,ij}^4} \right) \tilde{d}_{ij}^{q+1} \\
   & + \frac{\sigma_{1c,ij}}{2} \bar{\lambda}_{s,ij} \left\| W_{ij}^* \right\|^2 
   + \sigma_{2c,ij} \left\| W_{ij}^{*\frac{p+1}{2}} \right\|^2 
   + \bar{\sigma}_{3c,ij} \left\| W_{ij}^{*\frac{q+1}{2}} \right\|^2 \\
   & + \frac{1}{2} \bar{\delta}_{ij} 
   + \sigma_{1\theta,ij} \left\| \theta_{ij}^{*\frac{p+1}{2}} \right\|^2 
   + \bar{\sigma}_{2\theta,ij} \left\| \theta_{ij}^{*\frac{q+1}{2}} \right\|^2 \\
   & + \frac{3}{4} \left( \frac{\mu_{d,ij}}{\gamma_{d,ij}} \right)^{\frac{4}{3}} \bar{d}_{d,ij}^{\frac{4}{3}} 
   + \sigma_{1d,ij} \bar{d}_{ij}^{p+1} 
   + \sigma_{2d,ij} \bar{d}_{ij}^{q+1} 
   + z_{ij} z_{i(j+1)}.
\end{aligned}
\tag{D-2}
\end{equation}

The bounds on the design parameters specified in Lemma 10 (see \eqref{eq:lemma10_parameters}) ensures that the time derivative of the Lyapunov function in Step $j$ is bounded by, 

\begin{align*}
  & \dot{V}_{ij} \leq \dot{V}_{i(j-1)} - z_{ij} z_{i(j-1)}^r 
    - k_{p,ij} z_{ij}^{p+1} - k_{q,ij} z_{ij}^{q+1} \\
  & - \sigma_{2ca,ij} \left\| \tilde{W}_{c,ij}^{\frac{p+1}{2}} \right\|^2 
    - \sigma_{3ca,ij} \left\| \tilde{W}_{c,ij}^{\frac{q+1}{2}} \right\|^2  - \frac{\sigma_{2a,ij}}{2} \left\| \tilde{W}_{a,ij}^{\frac{p+1}{2}} \right\|^2 \\
  & 
    - \underline{\sigma}_{3a,ij} \left\| \tilde{W}_{a,ij}^{\frac{q+1}{2}} \right\|^2 - \frac{\sigma_{1\theta,ij}}{2} \left\| \tilde{\theta}_{ij}^{\frac{p+1}{2}} \right\|^2 
    - \underline{\sigma}_{2\theta,ij} \left\| \tilde{\theta}_{ij}^{\frac{q+1}{2}} \right\|^2 \\
  & - \frac{\sigma_{1d,ij}}{2} \tilde{d}_{ij}^{p+1} - \tilde{\sigma}_{2d,ij} \tilde{d}_{ij}^{q+1} 
    + C_{ij} + z_{ij} z_{i(j+1)} \\
  & = - \sum_{k=1}^j k_{p,ik} \left(z_{ik}^r\right)^{p+1} 
    - \sum_{k=1}^j k_{q,ik} \left(z_{ik}^r\right)^{q+1} \\
  & - \sum_{k=1}^j \sigma_{2ca,ik} \left\| \tilde{W}_{c,ik}^{\frac{p+1}{2}} \right\|^2 
    - \sum_{k=1}^j \sigma_{3ca,ik} \left\| \tilde{W}_{c,ik}^{\frac{q+1}{2}} \right\|^2 \\
  & - \frac{1}{2} \sum_{k=1}^j \sigma_{2a,ik} \left\| \tilde{W}_{a,ik}^{\frac{p+1}{2}} \right\|^2 
    - \sum_{k=1}^j \sigma_{3a,ik} \left\| \tilde{W}_{a,ik}^{\frac{q+1}{2}} \right\|^2 \\
  & - \frac{1}{2} \sum_{k=1}^j \sigma_{1\theta,ik} \left\| \tilde{\theta}_{ik}^{\frac{p+1}{2}} \right\|^2 - \sum_{k=1}^j \underline{\sigma}_{2\theta,ik} \left\| \tilde{\theta}_{ik}^{\frac{q+1}{2}} \right\|^2\\
   & 
    - \frac{1}{2} \sum_{k=1}^j \sigma_{1d,ik}^r \left( \tilde{d}_{ik}^r \right)^{p+1}  - \frac{1}{2} \sum_{k=1}^j \sigma_{2d,ik}^r \left( \tilde{d}_{ik}^r \right)^{q+1} \\
    &
    + \sum_{k=1}^j C_{ik} + z_{i(j+1)} z_{ij}.
\end{align*}

\bibliographystyle{unsrt}
\bibliography{MyReferences}

\end{document}